\documentclass[journal]{IEEEtran}

\usepackage{enumerate}
\usepackage{cite}
\usepackage{amsmath,amssymb,amsfonts}
\usepackage{algorithmic}
\usepackage[pdftex]{graphicx}
\usepackage{epstopdf}
\usepackage{color}
\usepackage{textcomp}
\usepackage{subfigure}
\usepackage{array,multirow,hhline}

\DeclareMathOperator*{\argmax}{arg\,max}
\begin{document}
%
\title{Using learning to control artificial avatars in human motor coordination tasks}
%
%
%

\author{Maria Lombardi, \IEEEmembership{Member, IEEE}, Davide Liuzza, and Mario di Bernardo, \IEEEmembership{Fellow, IEEE}
	\thanks{Submitted on 2nd November 2018}
	\thanks{Maria Lombardi is with the Department of Engineering Mathematics, University of Bristol, UK (e-mail: maria.lombardi@bristol.ac.uk) and with the Department of Electrical Engineering and Information Technology, University of Naples ``Federico II", Italy (e-mail: maria.lombardi@unina.it).}
	\thanks{Davide Liuzza is with ENEA Fusion and Nuclear Safety Department, Frascati (Rome), Italy (e-mail: davide.liuzza@enea.it).}
	\thanks{Mario di Bernardo is with the Department of Engineering Mathematics, University of Bristol, UK (e-mail: m.dibernardo@bristol.ac.uk) and with the Department of Electrical Engineering and Information Technology, University of Naples ``Federico II", Italy (e-mail: mario.dibernardo@unina.it).}}

\markboth{Journal of \LaTeX\ Class Files,~Vol.~xx, No.~x, month~year}%
{Shell \MakeLowercase{\textit{et al.}}: Bare Demo of IEEEtran.cls for IEEE Journals}

\maketitle

\begin{abstract}
Designing artificial cyber-agents able to interact with humans safely, smartly and in a natural way is a current open problem in control. Solving such an issue will allow the design of cyber-agents capable of co-operatively interacting with people in order to fulfil common joint tasks in a multitude of different applications. This is particularly relevant in the context of healthcare applications. Indeed, the use has been proposed of  artificial agents interacting and coordinating their movements with those of patients suffering from social or motor disorders. Specifically, it has been shown that an artificial agent moving its end effector with certain kinematic properties could provide innovative and efficient rehabilitation strategies for these patients. Moreover, it has also been shown that the level of motor coordination is enhanced if these kinematic properties are similar to those of the individual it is interacting with. In this paper we discuss, first, a new method based on Markov Chains to confer ``human motor characteristics'' on a virtual agent, so that it can coordinate its motion with that of a target individual while exhibiting specific kinematic properties. Then, we embed such synthetic model in a novel control architecture based on reinforcement learning to synthesise a cyber-agent able to mimic the behaviour of a specific human performing a joint motor task with one or more individuals.
\end{abstract}


\IEEEpeerreviewmaketitle

\section{Introduction}
\label{sec:introduction}
\IEEEPARstart{T}{he} number of  tasks involving people coordinating their movement with machines, avatars and robots is increasing at a rapid pace \cite{Iqbal2016,Atkeson2000,Shukla2012}. 
Investigating how to enable such artificial agents to co-operatively interact with humans is still an open problem. To achieve this ambitious goal, it is important to study how humans coordinate their motion with each other and then develop control strategies able to drive artificial agents in motor coordination tasks with them.

It has been observed that pairs or groups of humans performing a joint motor task often tend to intentionally or unintentionally synchronise their movement. Examples include  studies on people rocking chairs \cite{Richardson2007}, hands clapping \cite{Neda2000}, team rowing during a race \cite{Wing1995}, and synchronisation of respiratory rhythms \cite{Codrons2014}.
To study the emergence of interpersonal motor coordination  between two individuals, the ``mirror game'' was proposed in \cite{Noy2011} as a paradigmatic study case. In the mirror game, two players are asked to imitate each other's hand movement creating spontaneous coordinated motion (see Sec. \ref{sec:mirror_game_ims} for further details). 

In \cite{Slowinski2016},  it was shown from studying a large dataset of position and velocity time series of people playing the game that the motion of each individual exhibits unique kinematic features that can be summarised through  a time-invariant individual motor signature (IMS) captured by the distribution of the velocity profile exhibited during the motion (see Sec. \ref{sec:mirror_game_ims}). More importantly, it was also observed that when two players share a similar IMS then their movement coordination is significantly enhanced during the game.

Using this observation, it was suggested  and proved that the IMS can be used as a biomarker to assess patients suffering from social disorders such as schizophrenia as their IMS was found to significantly differ from those of healthy individuals \cite{Slowinski2017}. Moreover social disorders were found to have an impact on the ability of patients to coordinate their motor behaviour with that of others  \cite{DelMonte2013,Varlet2012} and that better coordination can be observed when the IMS of the patient is matched by that of the other player \cite{Slowinski2017} or by using social priming \cite{Raffard2015}. Similar observations were also made in the context of applications of social robotics to autism as for example in \cite{Wainer2014,Begum2016}.

Motivated by the potentially disruptive application of developing serious motor coordination exergames for the diagnosis and rehabilitation of social disorders, the problem was set in \cite{Zhai2015,Zhai2016} of designing cognitive architectures based on feedback control strategies able to drive an artificial avatar or robot to coordinate the motion of their end-effector with that of the arm or hand of a patient while exhibiting some desired IMS.

Previous approaches to design a control architecture to make an artificial avatar play the mirror game while exhibiting a desired IMS include those proposed in \cite{Zhai2016,Zhai2017} where model predictive control was used to solve a multi-objective control problem aimed at (1) tracking the motion of the end-effector of the human player (HP) or generating motion to lead the human player in the game; (2) exhibiting a desired IMS in the generated avatar motion.
As shown in detail in Sec. \ref{sec:control_architecture}, the solution proposed therein relies on a deterministic controller solving an optimal control problem on a receding horizon. The controller uses the nonlinear Haken-Kelso-Bunz oscillator \cite{Haken1985} to solve its model-based optimisation with the cost function being selected so as to minimise the mismatch between the positions of the end-effectors of the virtual player and the human player, while at the same time minimising the distance matching the velocity distribution of the virtual player motion with that of a pre-recorded human movement trajectory (treated as a reference IMS).

The key disadvantages of this approach are the deterministic nature of the controller and the model being used often leading to unnatural tracking behaviour and the need of pre-recorded human trajectories matching the desired properties of the reference IMS. Also, the dual nature of the control objective requires fine tuning of all control parameters in the cost function in order to achieve an acceptable balance between the two tracking errors rendering the use of those approaches unpractical in real interaction scenarios.

The goal of this paper is to use stochastic modelling and reinforcement learning algorithms to solve both of these problems.
 Specifically, we develop
\begin{enumerate}[i.]
	\item an {\em artificial IMS generator} based on data-driven stochastic Markov chain models that is able to learn the IMS of a target individual and then autonomously generate reference motion endowed with that IMS, removing the need of any pre-recorded signal;
	\item a new cognitive architecture that exploiting such an IMS generator and using a reinforcement learning (RL) algorithm is  able to drive the artificial agent (or Cyberplayer -- CP -- in the rest of this paper) to play the mirror game with another player (human or artificial) while at the same time exhibiting a desired IMS, overcoming all the limitations of previous control solutions.
\end{enumerate}

Also, we solve the problem of the RL requiring a large amount of training data by synthesising one or more ``virtual trainers'' (VTs) able to generate as much synthetic data as needed during the training phase. Such VTs are driven by the model-based control architecture proposed in \cite{Zhai2016} aptly modified by embedding in its core the IMS generator presented in this paper, see Sec. \ref{sec:modelling} for further details.
In so doing, the Cyberplayer is synthesised as an artificial agent able to train itself by observing the VTs playing the mirror game with each other or with other human players. After presenting the derivation of the IMS generator and the design of the RL-based cognitive architecture, we use numerical simulations and real experiments where the CP is set to play with human players to validate and confirm the effectiveness of the proposed approach. All the experiments are carried out using CHRONOS, the experimental set-up to study motor coordination in dyadic or group sessions which was first presented in \cite{Alderisio2017b,Chronos} and is summarised in Sec. \ref{sec:validation} for the sake of completeness.

We wish to emphasise that the problem addressed in this paper is an instance of the larger class of problems involving robots or artificial avatars interacting with a \emph{human-in-the-loop}, which are currently the subject of much ongoing research, see for example \cite{Iqbal2016,Atkeson2000,Shukla2012}. 
The rest of the paper is organised as follows. After giving the background in Sec. \ref{sec:background},  we derive a data-driven model in Sec. \ref{sec:modelling} able to endow the VT with human kinematic features and use it as part of a control architecture that allows VT to play dyadic sessions of the mirror game with a human player as shown in Sec. \ref{sec:vp_modelling}. In Sec. \ref{sec:cp_modelling} we synthesise a cyber player through techniques of artificial intelligence. All these methods are validated in Sec. \ref{sec:validation} showing their effectiveness before conclusions are drawn in Sec. \ref{sec:conclusions}.


\section{Background}
\label{sec:background}

\subsection{Mirror game}
\label{sec:mirror_game_ims}
The mirror game is a serious game in which two people imitate each other's movements creating fascinating choreographies with their bodies. Usually played by musicians, dancers and actors, the mirror game has become a powerful paradigm to study the complex phenomenon of human motor coordination \cite{Noy2011}. In \cite{Noy2011} a simplest one-dimensional instance of the game was proposed as a  paradigmatic case of study where  two people mirror each other's movements by oscillating two handles horizontally sideways. 

The game can be played in three different conditions:
\begin{enumerate}
	\item Leader - Follower (LF): in this condition one player designated as follower tries to imitate as better as s/he can the trajectory of the other player designated as leader;
	\item Joint - Improvisation (JI): in which two players play together coordinating their movements without explicit designation of roles;
	\item Solo Condition (SC):  in which each player is asked to generate interesting motion in isolation without interacting with the other player. 
\end{enumerate}

\subsection{Cognitive architecture}
\label{sec:control_architecture}

\begin{figure}[tb]
	\framebox{\parbox{0.97\columnwidth}{
			\centering
			\includegraphics[width=0.95\columnwidth]{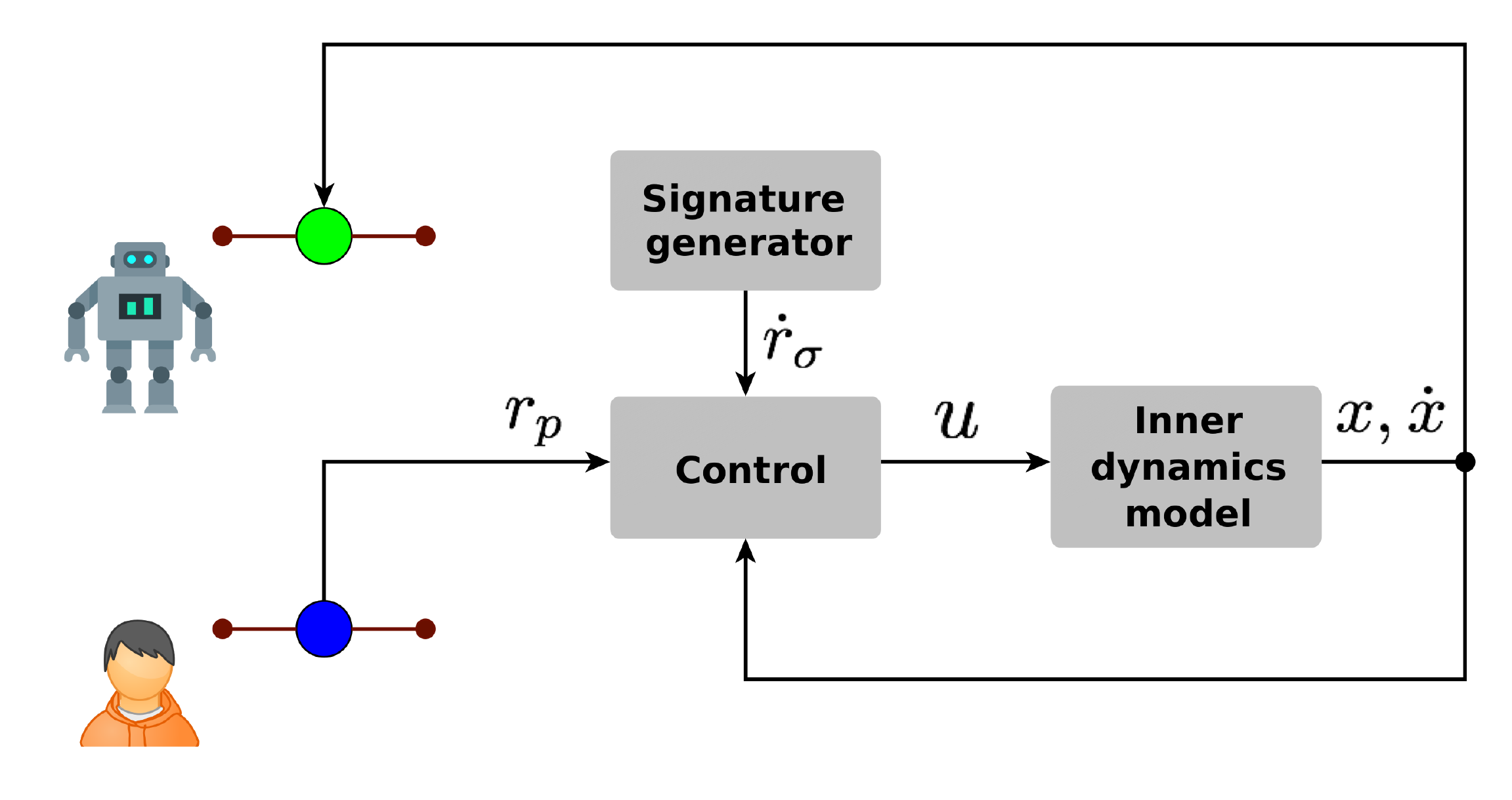}}}
	\caption{Schematic diagram of the model-based cognitive architecture proposed in \cite{Zhai2016} that allows an artificial agent to play sessions of the mirror game with another player. While the HP is playing the mirror game moving a spherical handle (in blue) from left to right, his/her position $r_p$ is sensed and given as a reference input to the model-based control algorithm together with a prerecorded signal associated to the desired kinematic signature $\dot{r}_\sigma$. The controller then drives an inner dynamic model of the end-effector dynamics to generate the position $x$ and the velocity $\dot{x}$ that move the VT's handle (in green).}
	\label{fig:control_architecture}
\end{figure}

A cognitive architecture was proposed in \cite{Zhai2016, Zhai2017} to drive a virtual agent in playing the mirror game with a human player while exhibiting some reference IMS. The architecture is shown in Fig. \ref{fig:control_architecture} and is mainly composed of two parts:
\begin{itemize}
	\item \textit{an inner dynamics model} representing how the artificial agent moves in the absence of any interaction with the HP. This was modelled using a nonlinear Haken-Kelso-Bunz oscillator (HKB) \cite{Haken1985, Zhai2016, Zhai2017} which is often used in the human movement modelling literature;
	\item \textit{a control strategy} that generates the movement of the agent in response to that of the human player while exhibiting the reference IMS (velocity distribution) provided by a ``signature generator'' block. In \cite{Zhai2015, Zhai2016, Zhai2017} an optimal controller was proposed in order to minimise the difference in position between the HP and the end-effector of the artificial agent. At the same time the cost function adopted therein took into account the error in velocity between the agent's motion and a reference IMS which was pre-recorded during off-line human-human game trials.
\end{itemize}

More specifically, in \cite{Zhai2016,Zhai2017}  the motion of the virtual agent is modelled as a controlled nonlinear HKB oscillator of the form:
\begin{equation}
\ddot{x} + \left( \alpha x^2 + \beta \dot{x}^2 - \gamma \right) \dot{x} + \omega^2 x = u,
\end{equation}
where  $x, \dot{x}$ and $\ddot{x}$ are the position, velocity and acceleration of its end-effector,  $u$ is the control input, $\alpha, \beta, \gamma$ are parameters characterising the nonlinear damping term while $\omega$ is the natural oscillation frequency of the generated motion when $u$ is set to zero. 

According to \cite{Zhai2016, Zhai2017}, the control input $u$, is chosen as a function of the movement of the HP in order to minimise over the time interval $[t_k, t_{k+1}]$ the following cost function
\begin{multline}
\min\limits_{u} J\left( t_k \right) = \frac{1}{2} \theta_p \left( x\left(t_{k+1}\right) - r_p\left(t_{k+1}\right) \right)^2 + \\
\frac{1}{2} \int_{t_k}^{t_{k+1}}{\left(1-\theta_p\right) \left(\dot{x}\left(\tau\right) - \dot{r}_\sigma\left(\tau\right) \right)^2 + \eta u\left(\tau\right)^2 d\tau},
\label{eq:controller}
\end{multline}
where $r_p$ is the position time series measured from the HP end-effector, $\dot{r}_\sigma$ is the reference signal corresponding to the desired motor signature, $\eta$ is a positive weight assigned to the minimisation of the control energy. The constant parameter $\theta_p \in [0,1]$ is used to determine how much the agent weighs the motion of the human player so that if $\theta_p=0$ the agent acts as a leader (completely ignoring the motion of the other player) while if $\theta_p=1$ as a perfect follower. Any value of $\theta_p \in ]0,1[$ will make the agent's motion exhibit a compromise between tracking the HP motion and the reference velocity of the target IMS it has to exhibit, allowing to implement different types of leader/follower behaviour.

The proposed control architecture requires two different reference signals, one is the position signal recorded by the HP, the other is a velocity profile that represents the desired IMS for the artificial agent. The main drawback in the solution adopted in \cite{Zhai2016, Zhai2017} is that the IMS is collected off-line, recording several human players performing sessions of the mirror game in solo condition. The use of a pre-recorded signature makes the motion of the agent less natural and less variable since the reference is always the same for each session of the game making its use as a leader over several trials repetitive and not engaging for the human player it is playing against. 

In what follows, to overcome these problems we will take a different approach and use Markov chains to obtain data driven models of HPs performing solo sessions of the mirror game. These models will be integrated into the control scheme of Fig. \ref{fig:control_architecture} as stochastic signature generators so as to endow the artificial agent with greater flexibility and variability between one session and the other. This approach will the be used in the rest of the paper to synthesise an artificial, or virtual trainer (VT) able to generate synthetic data to train a new type of artificial agent (or Cyberplayer) driven by a reinforcement learning strategy rather than deterministic control approaches. 

\subsection{Reinforcement Learning}
\label{sec:reinforcement_learning}
Reinforcement learning (RL) is a machine learning technique in which an agent learns how to behave in an external environment through a trial-and-error approach and looking at its successes and failures \cite{Russel2003, Sutton1998}. RL techniques are well suited when agents are formally modelled as a Markov Decision Process (MDP), composed of the quadruple $\left\langle X, U, f, \rho \right\rangle$ where $X$ is the set of all possible states in which the environment can be, $U$ is the set of all possible actions that the agent can take (also termed as action-space), $f:X \times U \times X \rightarrow \left[0,1\right]$ is the state transition probability function and $\rho:X \times U \times X \rightarrow \mathbb{R}$ is the reward function (see \cite{White1989} for more details on MDP).
The learning process is made up of the following sequential steps:
\begin{enumerate}
	\item the agent observes the environment (process and measurements of interest) and takes an action $u$ in the set of all possible actions $U$, which causes the environment to transit into a new state;
	\item the agent records the new state of the environment $\textbf{x} \in X$ following its action;
	\item the agent receives a scalar reward $r$ that measures how good taking that action in the previous state has been;
	\item according to the reward, the agent changes its policy $\pi: X \rightarrow U$ that maps each environment state $\textbf{x}$ to an action $u$;
	\item the agent aims at maximising the sum of all rewards obtained along all the interactions.
\end{enumerate}
Solving a problem of RL mathematically is equivalent to solving a system of $N_x \times N_u$ non linear equations, where $N_u$ is the number of all possible actions and $N_x$ is the number of all possible states. Due to the complexity of such a problem, in particular for huge state space dimensions, several iterative approaches can be used, for which it has been proved that the policy followed by the agent converges to the optimal policy (see \cite{Sutton1998} for further details). In this work, we use a Temporal Difference Learning approach, in particular the Q-learning algorithm since it does not require the model of the environment and it operates completely online estimating the reward and the state value.

\section{IMS Generator Design}
\label{sec:modelling}
Next Markov chains are used to model human movement in solo condition and design an IMS generator. A Markov Chain (MC) is a well-known stochastic model useful to describe randomly changing systems with a finite number of states \cite{Rabiner1989}. It consists of a finite set of possible states of the system, a set of transitions between any two states with their corresponding probabilities, and a set of possible observations or outputs. 

Denoting by $s_k = i$ with $i \in [1,\ldots,N]$ that the system is in state $i$ at time $k$, a Markov chain is fully characterised by:
\begin{itemize}
	\item an initial state $s_0$;
	\item a transition matrix $A := [a_{ij}]$ where $a_{ij} := P\left(s_{k+1} = j | s_k = i \right)$ is the probability of being in state $j$ at time $k+1$ having been in state $i$ at time $k$.
\end{itemize}

The procedure to construct a MC-based model consists of the following three steps:
\begin{enumerate}
	\item \textit{Data collection and preprocessing}: input movement data recorded from a human player performing the mirror game in solo condition are preprocessed through short-time Fourier Transform (STFT) and  Vector Quantization (VQ) \cite{Ahmed2012}.
	\item \textit{Markov model training}: the preprocessed data are used to define the coefficients of the transition matrix $A$ of the Markov chain.
	\item \textit{Synthetic data movement generation}: the resulting Markov chain is used to  generate new synthetic movement data sharing the same IMS as that of the input data.
\end{enumerate}
Next we describe in greater detail how each of these steps was performed.

\subsection{Step 1: Data collection and preprocessing}
\begin{figure}[thpb]
	\framebox{\parbox{0.97\columnwidth}{
			\centering
			\includegraphics[trim={1.5cm 0 0 0}, clip, width=\columnwidth]{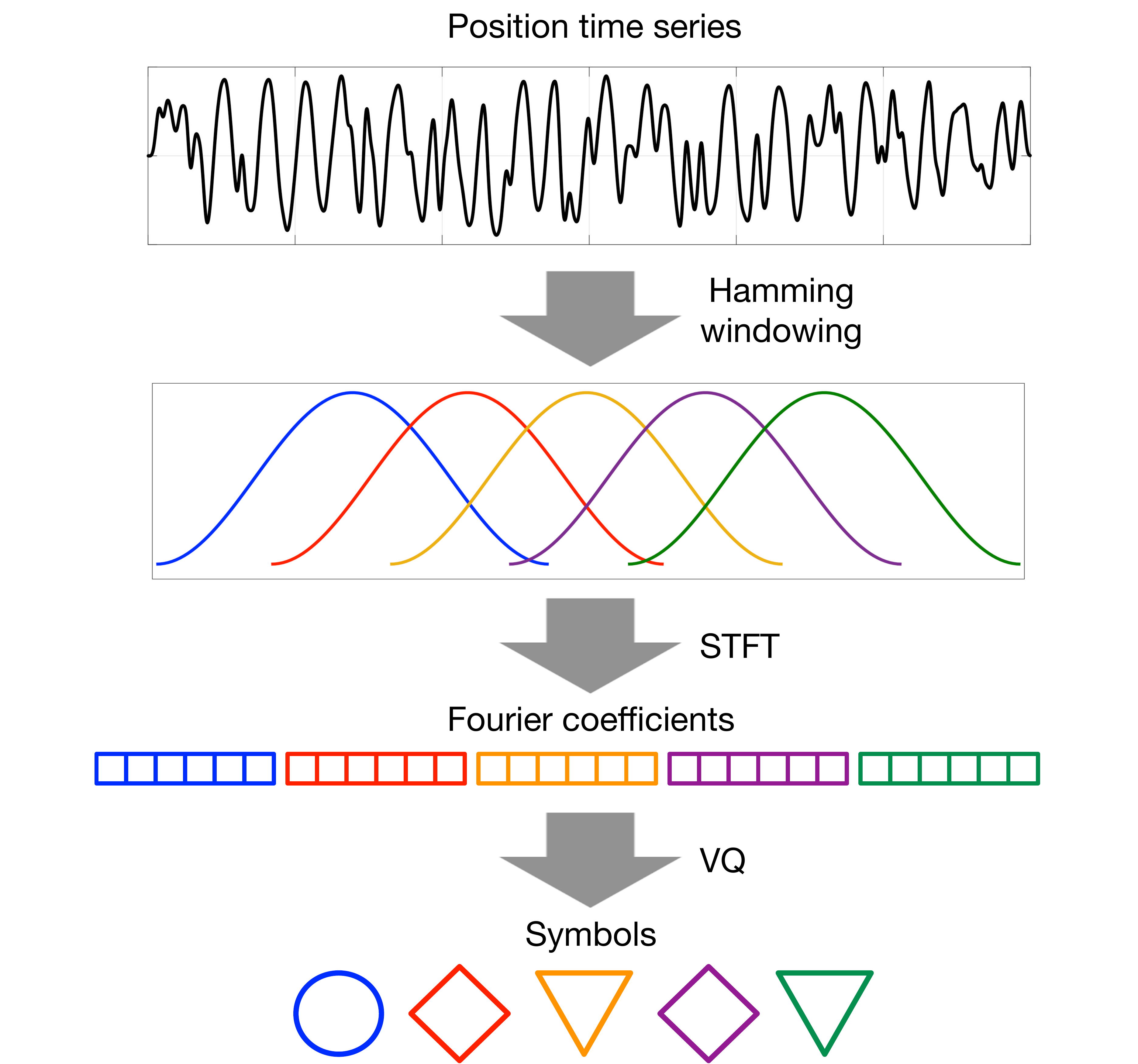}}}
	\caption{Illustration of the MC modelling process. The recorded position time series is windowed through the Hamming window. A feature vector is extracted from each window through the STFT and then quantized and converted into a symbol sequence by means of a vector quantizer. Different colours are used to highlight the preprocessing steps for different frames of the signal.}
	\label{fig:mc_modelling}
\end{figure}

The sampled input signal is first partitioned in a discrete set of frames using a Hamming window \cite{Heinzel2002} of a certain width (see Fig. \ref{fig:mc_modelling}). To prevent loss of information and in order to minimise the distortion of the signal, two consecutive windows are overlapped of $\frac{3}{4}$ of the window's width. In so doing the sum of the sequence of windows is a resulting flat-top window \cite{Heinzel2002}.

Each windowed data segment undergoes a Short Time Fourier Transform (STFT) so that a vector of Fourier transform's coefficients, or ``feature vector'', can be associated to it. 

At the end of the process a finite set of feature vectors is obtained that is processed through a vector quantizer in order to  get a finite set of symbols. The vector quantizer maps each feature vector to one of the $N$ prototype vectors contained in its code-book (for further details see \cite{Russel2003}). In this work, the code-book is generated by means of Lloyds algorithm \cite{Russel2003}. The latter is an iterative algorithm that continuously partitions Euclidean space into $N$ convex cells (also known as Voronoi cells) in order to have each input vector as close as possible to the centroid of one of the cells. The indexes of the array built with the $N$ prototype vectors (integers from $0$ to $N-1$), are used as symbols in the output of the discrete MC.

\subsection{Step 2: Markov model training}
Once the input data has been transformed into a finite set of symbols, the transition probability between one symbol and any other can be evaluated by looking at the symbol frequency in the data string. This allows to identify the coefficients of the transition matrix $A$ and so to build a MC model in which each state corresponds to a code-book symbol (the model has as many states as symbols). 
Note that the choice of having a Markov chain with as many states as symbols was found to be sufficient (and easier to train) for the problem of interest. In particular we found that choosing a codebook with $N$ prototyping vectors allows us to capture correctly the movement variability detected in the time series of the human player's movement.

In this paper, the transition coefficients are evaluated over an extensive set of experimental data through the Baum-Welch algorithm \cite{Rabiner1989}, which is essentially based on a frequential approach, and finds the maximum likelihood estimate of the coefficients given a set of observed feature vectors.

\subsection{Step 3: Synthetic data movement generation}
Because of its stochastic nature, the MC built in the previous step can generate random sequences of symbols according to the probabilities included in the transition matrix $A$. In order to have a position signal over time, the sequence of symbols needs to be reconverted to a continuous signal through reverse post-processing. To this aim, the generated sequence of symbols is de-quantized with the same code-book used in the forward pre-processing so that each symbol is mapped back onto a prototype feature vector. Then, an inverse STFT is performed of each feature vector and concatenated using the overlap-add (OLA) method \cite{Rabiner1975}. The main advantage of this method is the possibility of reconstructing from the generated symbols a smooth position time series, removing any discontinuity caused by simply concatenating two random symbols.


\begin{figure*}[thpb]
	\framebox{\parbox{0.98\textwidth}{
			\centering
			\vspace{5pt}
			\includegraphics[width=0.85\textwidth]{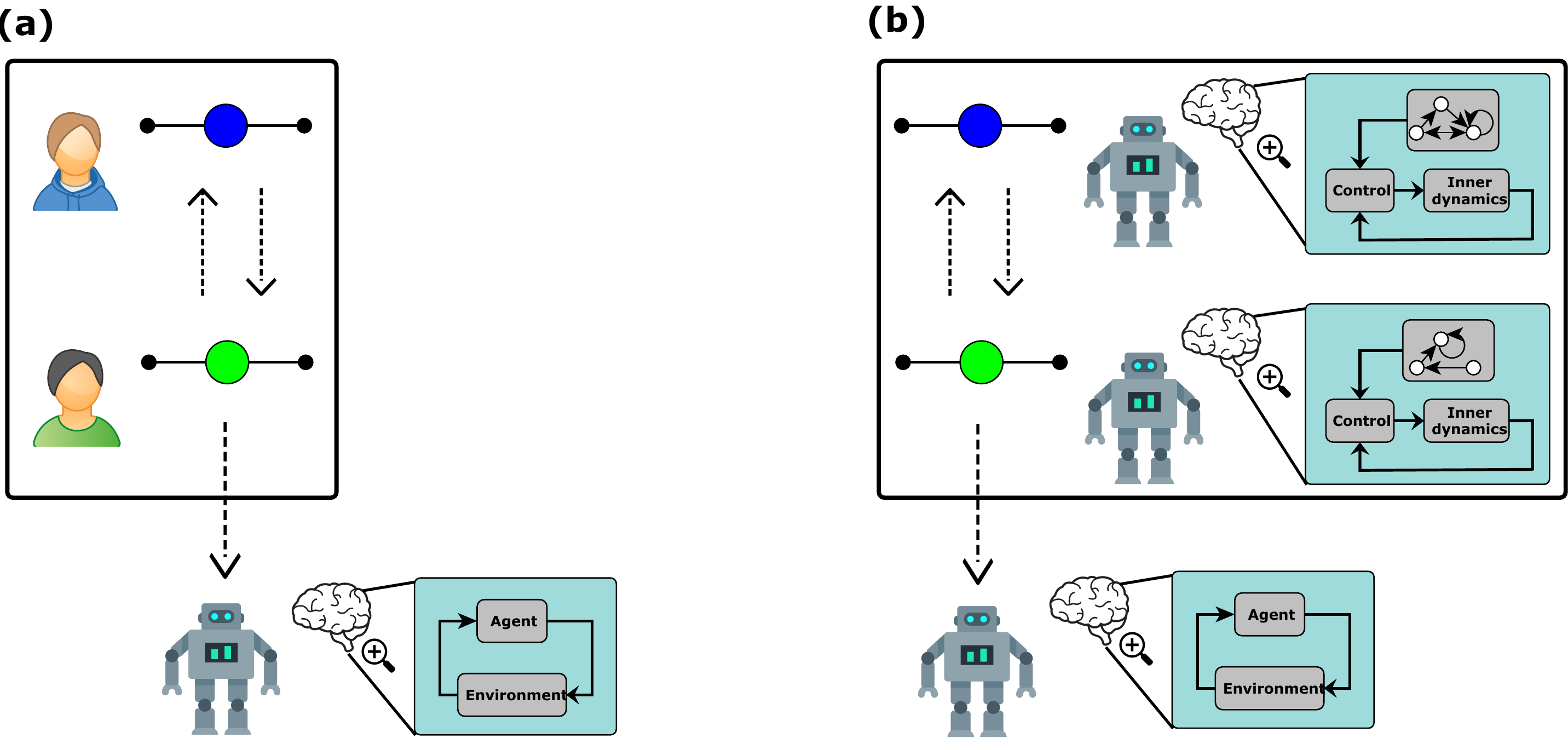}}}
			\vspace{5pt}
	\caption{Schematic of the possible training stage for the CP. \textbf{(a)} The CP is trained by observing a human player playing sessions of the mirror game in a leader-follower configuration. \textbf{(b)}  The two HPs are substituted with two model-based virtual trainers playing against each other. Each of the VT is driven by a model-based controller endowed with an IMS Generator synthesised using data acquired during a much smaller number of solo sessions of each of the HPs.}
	\label{fig:2players_system}
\end{figure*}

\section{Control synthesis}
\subsection{MC in the loop}
\label{sec:vp_modelling}
As a first step, we embed the MC model developed above to replace the ``signature generator'' block in the control schematic depicted in Fig. \ref{fig:control_architecture} and described in Sec. \ref{sec:control_architecture}. By using the MC model trained on data acquired from a target HP, the artificial agent driven by the MC endowed control architecture will be able to play the mirror game with another player while exhibiting the IMS of the target HP the MC was modelled upon.
The quality of the tracking and of the IMS exhibited by the agent will still depend upon the many parameters of the optimal control approach described in Sec. \ref{sec:background} and \cite{Zhai2017} that require careful off-line tuning and numerous trial-and-error experiments. To overcome this problem we take next a radically different approach based on the use of reinforcement learning techniques which will lead to a fully data-driven control algorithm.

\subsection{A reinforcement learning approach}
\label{sec:cp_modelling}

\begin{figure*}[tb]
	\framebox{\parbox{0.98\textwidth}{
			\centering
			\vspace{8pt}
			\includegraphics[width=0.7\textwidth]{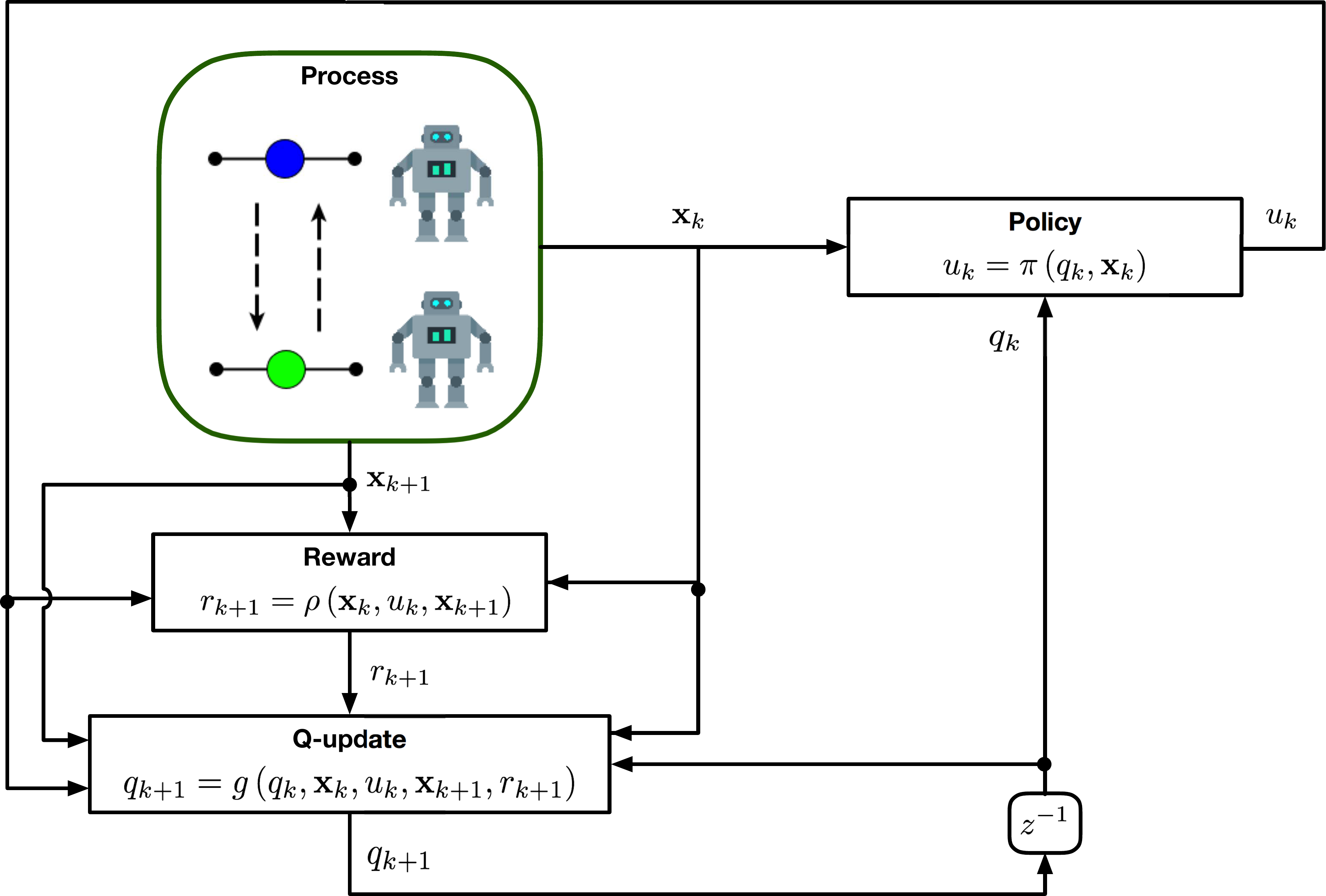}
			\vspace{8pt}}}
	\caption{Block diagram summarising the reinforcement learning algorithm used to drive the Cyberplayer. The Q-learning controller chooses a control input $u$ according to the current Q-table and process state. The latter evolves in a new state  $\mathbf{x}$ used to evaluate the instant reward and update consequently the Q-table. With the updated table, the controller chooses the next action taken by the CP.}
	\label{fig:RL_block_diagram}
\end{figure*}

The goal of the reinforcement learning approach is to develop an artificial agent (or Cyber-Player (CP))   able to play dyadic sessions of the mirror game in any desired condition, acting as a leader or a follower, while exhibiting the IMS of a target human player. Note that the aim is for the CP to emulate the way in which a specific human player would interact with the another player in a mirror game session with all of his/her ``human imperfections" rather than achieving perfect tracking of the partner position.

To achieve this goal, the reinforcement learning algorithm needs to be provided with data from leader-follower sessions of the mirror game so that the CP could learn to mimic the behaviour of the target player in these sessions. Specifically, particularising the learning process described in Sec. \ref{sec:reinforcement_learning} to our specific case, we consider as system's state $\mathbf{x} := \left[x, \dot{x}, x_p, \dot{x}_p \right]$, where $\left[x, \dot{x}\right]$ are position and velocity of the CP, while $\left[x_p, \dot{x}_p\right]$ are position and velocity of the partner player. The reward function as $\rho := -\left(x-x_t\right)^2 - 0.1\left(\dot{x}-\dot{x}_t\right)^2 - \eta u^2 $ where  $\left[x_t, \dot{x}_t\right]$ are position and velocity of the target player (while playing with the other), $\eta$ is a positive weight for the minimisation of the control energy $u$. The choice  of such reward function is motivated by the fact that we want the CP to mimic the behaviour of the emulating player, thus behaving as him/her synthetic avatar.
To maximise this function in any situation,  the  cyber  player  selects  the  best  action  it can actually perform from a set of finite action values. In this context, the action space consists of values of acceleration of the end-effector of the cyber player. Such values can be positive or negative according to the cyber player movements (towards the right or towards the left). To estimate the action space and the range of acceleration values to be discretized, experiments were performed making two human players play sessions of the mirror game in a leader follower configuration. Analysing the data, a typical range was identified of acceleration values assumed by the human players. Using these values, it was found heuristically that a discretization using $9$ different action values represent a good compromise between the quality of the  resulting motion and the learning time.

To implement reinforcement learning we use the Q-learning algorithm. As anticipated in Sec. \ref{sec:reinforcement_learning}, this is an iterative approach, in which the CP adapts its behaviour according to the measures it receives from the two players (leader and follower) and trying to find the best action that it can perform for each state to emulate the target player. As a first step we define a matrix $Q := [q_{ij}]$ where the states are listed on the rows and the actions on the columns. Each element $q_{ij}$ is the ``value" given to the corresponding state-action pair, also termed q-value. At the beginning of the learning process, the matrix $Q$ is initialised with random values. Then, each iteration is structured as follows (see Fig. \ref{fig:RL_block_diagram}):
\begin{itemize}
	\item the CP observes the state $\textbf{x}_k$ (the subscript $k$ denotes the sampled value of the state at time instant $k$) and the target player state needed to evaluate the reward function;
	\item the CP chooses an action $u_k$ at a time instant $k$ according to a policy rule $\pi$. In this work we use an $\epsilon$-greedy policy \cite{Sutton1998}. Specifically, the CP takes the best known action, i.e, the action with the highest q-value (exploitation) with $\left(1-\epsilon\right)$ probability, whereas with $\epsilon$ probability it takes a random action (exploration). The value $\epsilon$ follows a monotonic decreasing function, since as time increases the exploration phase is replaced by the exploitation phase;
	\item the CP evolves in a new state at time $k+1$ and observes the state $\textbf{x}_{k+1}$ and the reward $r_{k+1} = \rho\left(\textbf{x}_k, u_k, \textbf{x}_{k+1}\right)$ following the action $u_k$ taken at the previous time $k$. For the sake of brevity, we will simply denote the obtained reward as $r_{k+1}$, omitting the dependence from the state and the actions;
	\item according to the reward received, the CP updates the value of the entry of the matrix $Q$ corresponding to the pair $\left(\mathbf{x}_k,u_k\right)$ following the rule:
	\begin{multline}
	q_{k+1}\left(\textbf{x}_{k}, u_{k}\right) = q_k\left(\textbf{x}_{k}, u_{k}\right) + \alpha\Biggl[ r_{k+1} + \\
	+ \gamma \argmax_{u_{k+1} \in U}q_k\left(\textbf{x}_{k+1}, u_{k+1}\right)- q_k\left(\textbf{x}_{k}, u_{k}\right) \Biggr],
	\end{multline}
	where $\alpha$ is the learning rate and $\gamma$ is a discount factor;
	\item a new iteration is performed until convergence is achieved.
\end{itemize}

\subsection{Training}
Once the learning algorithm has been defined, training trials must be arranged so that the algorithm can be fed with a large enough dataset for it to converge towards a viable control solution.
As the goal of the RL algorithm is to make the CP able to play the game while mimicking the kinematic features (IMS) of a target human player, position and velocity time series of such a human player should be ideally collected during several live sessions of the mirror game. As learning typically requires a large dataset, real data from human players might be difficult to collect. 

To overcome this problem, here we propose a practical way of training the CP by using synthetic data generated by two ``virtual trainers'' playing the mirror game against each other or with other human players. In this set-up, each VT is driven by the control architecture shown in Fig. \ref{fig:2players_system}, embedding as signature generator blocks the IMS Generator synthesised, as described in Sec. \ref{sec:vp_modelling}, to mimic the behaviour of different human players. 
Note that to synthesise the IMS Generator only a small dataset obtained during sessions of game where the HP plays in solo condition are needed. By embedding the IMS Generator in a model-based VT we can then generate much larger synthetic datasets from VTs playing dyadic session of the game that can be used to train the CP. Later in Sec. \ref{sec:validation} we will show that this training approach is able to endow the CP with a behaviour close to that of the human player it is learning to imitate.

\section{Validation}
\label{sec:validation}

\begin{figure}[tb]
	\framebox{\parbox{0.98\columnwidth}{
			\centering
			\includegraphics[width=0.97\columnwidth]{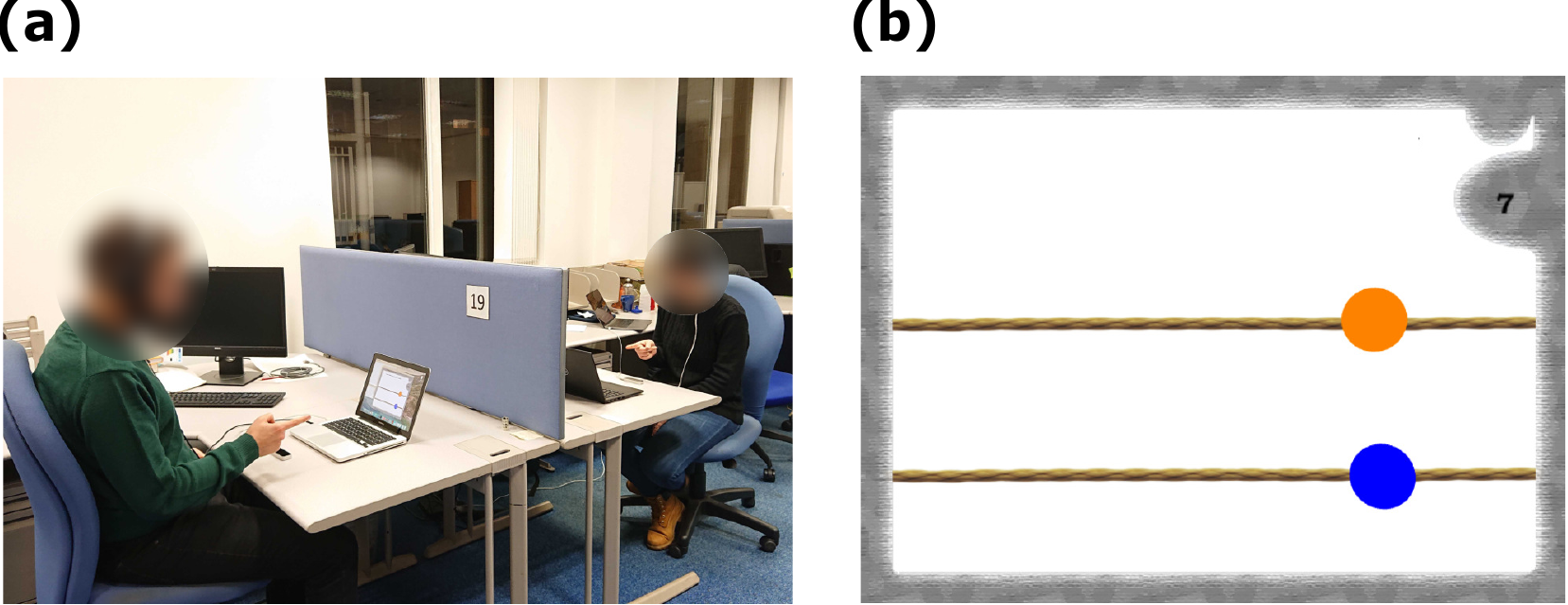}}}
	\caption{Experimental setup of the Mirror game using CHRONOS. \textbf{(a)} Two humans performing a dyadic trial moving their index fingers over a Leap motion controller. \textbf{(b)} User interface seen by the human players on their respective screen; each solid circle representing the motion of one of the two players.}
	\label{fig:mirrorgame_setup}
\end{figure}

To validate our methodology, we carried out experiments with the following features.
\begin{itemize}
	\item \textit{Participants}: a total of $6$ people took part in the experiments ($4$ females and $2$ males). Experiments were divided into two different sessions: solo experiments and dyadic experiments. All participants contributed to the validation of the MC modelling approach performing solo experiments, whereas $5$ participants out $6$ participated to the dyadic experiments for the validation of the RL-based CP.
	All participants were right-handed and none of them had physical or mental disabilities. All of them participated voluntarily, signing an informed consent form in accordance with the Declaration of Helsinki. The players were assigned numbers from $1$ to $6$.
	\item \textit{Experimental task}: each participant was asked to carry out $30$ different trials of $30$ seconds of the mirror game in solo condition. The given instruction was to oscillate the index finger of their preferred hand in a spontaneous way generating interesting sideways motion as smoothly as possible over a position sensor.
	$4$ participants were then asked to perform $8$ HP-HP trials of the mirror game each lasting $60$ seconds in leader-follower condition with the $5$-th participant acting as a follower. Then, for each pair, the $5$-th participant was replaced by the CP trained after him/her. Once again, $8$ HP-CP trials lasting $60$ seconds each were performed in the leader-follower condition for each pair. The given instruction to the HP in this case was to try to coordinate their own motion with that of the virtual partner.
	\item \textit{Experimental set-up}: experiments were performed through CHRONOS, a software tool recently developed  to study movement coordination which was presented in \cite{Alderisio2017b, Chronos}. CHRONOS is a software platform running over a dedicated WLAN network. It allows to perform solo and dyadic trials of the mirror game by making players see their own movement and that of the other players on the screen of their PC, removing any social interaction through visual or auditory coupling between the participants. CHRONOS also allows the deployment of one or more CPs to replace HPs.
	In our set-up, leap motion controllers \cite{Leapmotion} were used as sensors to acquire the position of the players' fingers (see Fig. \ref{fig:mirrorgame_setup}). To avoid delays in the communication between the laptops, a sampling frequency of $10$ Hz was found to be low enough to guarantee good communication performance while acquiring a sufficiently rich time series to be analysed. An upsampling to $100$ Hz was performed a posteriori before the analysis.
\end{itemize}

\subsection{Validation of the IMS Generator}
\label{sec:mc_validation}
To validate the ability of the IMS Generator based on the MC model synthesis process presented in Sec. \ref{sec:modelling} of generating motion characterised by the IMS of a target human player, we compared the velocity distribution of the motion of a HP recorded during a solo condition with that of the motion generated by the IMS Generator trained after him/her.
To compare the PDFs of the two velocity signals we used the Earth Mover Distance (EMD) as a viable metric \cite{Borg2012}. In the case of univariate probability distributions, the EMD is given by the area of the difference between their Cumulative Distribution Functions (CDF). Formally the distance between the velocity distribution of the human player, say $PDF_{HP}$, and that generated by the IMS Generator, say $PDF_{IMSG}$, is given by  
	\begin{multline}
	EMD(PDF_{HP}\left(z\right), PDF_{IMSG}\left(z\right)) = \\
	\int_{Z} \left| CDF_{HP}(z) - CDF_{IMSG}(z \right)| \, dz.
	\end{multline}
	Since a normalised EMD is used,  EMD equals 0 means that the two velocity profiles are identical and so perfect IMS matching is achieved while the EMD becomes increasingly larger as the overlap between the two IMS is completely absent.
	
Through multidimensional scaling (MDS) \cite{Borg2012,Slowinski2017}, it is possible to represent the player's velocity profiles as points in an abstract geometric space, called ``similarity space". Points corresponding to different trials of the same players can be encircled  by an ellipse defining a ``characteristic region'' for each player. The MDS is a technique that allows to reduce the dimensionality of the data, preserving as much information as possible (see \cite{Borg2012} for more details). The similarity space was highlighted as a valuable tool to analyse IMS in movement data in \cite{Slowinski2017}. Specifically,  the Euclidean distance between two points in the similarity space is a good approximation of the EMD between the corresponding PDFs; the closer the points are, the more similar are the corresponding IMS they associated with.

To show that the trajectories generated by the IMS Generator have the same features of a human movement, as described in \cite{Noy2011} we also computed  the \textit{kurtosis} and \textit{skewness} of the motion signals. 
In particular, given a generic smooth curve $f\left(t\right)$ with mean $\mu$ and standard deviation $\sigma$ defined on the interval $T = \left[t_1, t_2\right]$, its skewness and kurtosis are defined respectively as
\begin{equation}
s = \dfrac{1}{\sigma^{\frac{3}{2}}} \int_{t_1}^{t_2} \left( t-\mu \right)^3 f \left( t \right) \, dt,
\end{equation}
\begin{equation}
\kappa = \dfrac{1}{\sigma^{2}} \int_{t_1}^{t_2} \left( t-\mu \right)^4 f \left( t \right) \, dt.
\end{equation}
Roughly speaking, skewness indicates the asymmetry in acceleration and deceleration, whereas kurtosis provides information about the uniformity of the maximal velocity. Low kurtosis means that an object is quickly accelerating and decelerating and keeps constant the velocity in between, conversely high kurtosis means that the object is accelerating and decelerating slowly keeping the maximal velocity for a short time \cite{Slowinski2016}.

\begin{figure}[thpb]
	\framebox{\parbox{0.97\columnwidth}{
			\centering
			\vspace{5pt}
			\includegraphics[width=\columnwidth]{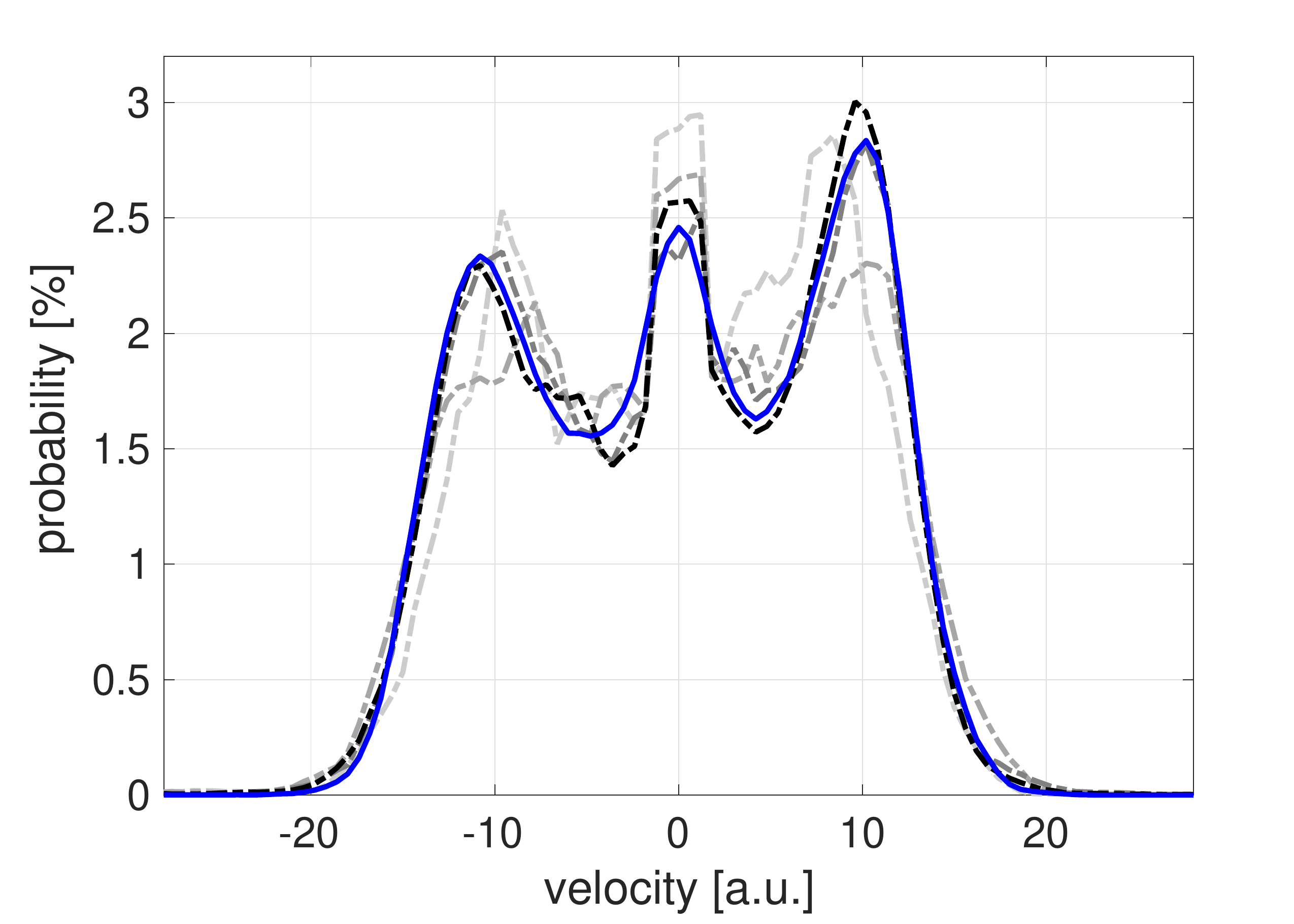}}}
	\caption{Comparison between the IMS of the human player in solo condition (in blue) with those generated by an IMS generator using Markov Chains with different cardinalities of the codebook (dashed lines spanning from light grey to black respectively for codebook of 32, 64, 128 and 256 symbols.}
	\label{fig:codebook_mc2}
\end{figure}

In our experiments, each position time series, also termed as a ``trial", was captured at a sampling rate of $10$ Hz. It was then interpolated to $100$ Hz and windowed with a $60$ samples long Hamming window with an overlapping of $45$ samples. 
The vector quantizer was built choosing the number of levels characterising its codebook in a heuristic manner. In particular, we tested different numbers of levels from $32$ to $256$ in order to find a good trade-off between the distortion error level and the number $N$ of levels required. At the end of the process, the vector quantizer was chosen to have $256$ levels (each level corresponds to a symbol). Fig. \ref{fig:codebook_mc2} shows a comparison between the IMS provided by the MC model and that of the human player in solo condition with different cardinalities of the codebook.

At the end six IMS Generators with $256$ states each were derived, each able to capture and reproduce the IMS of one of the players involved in the validation experiments . To assess the quality of the signatures produced by each generator, a total of $30$ new synthetic motion trajectories were generated and compared against those of the corresponding human players they were trying to mimic.

\begin{figure*}[thpb]
		\framebox{\parbox{0.97\textwidth}{
			\centering
			\includegraphics[width=0.44\textwidth]{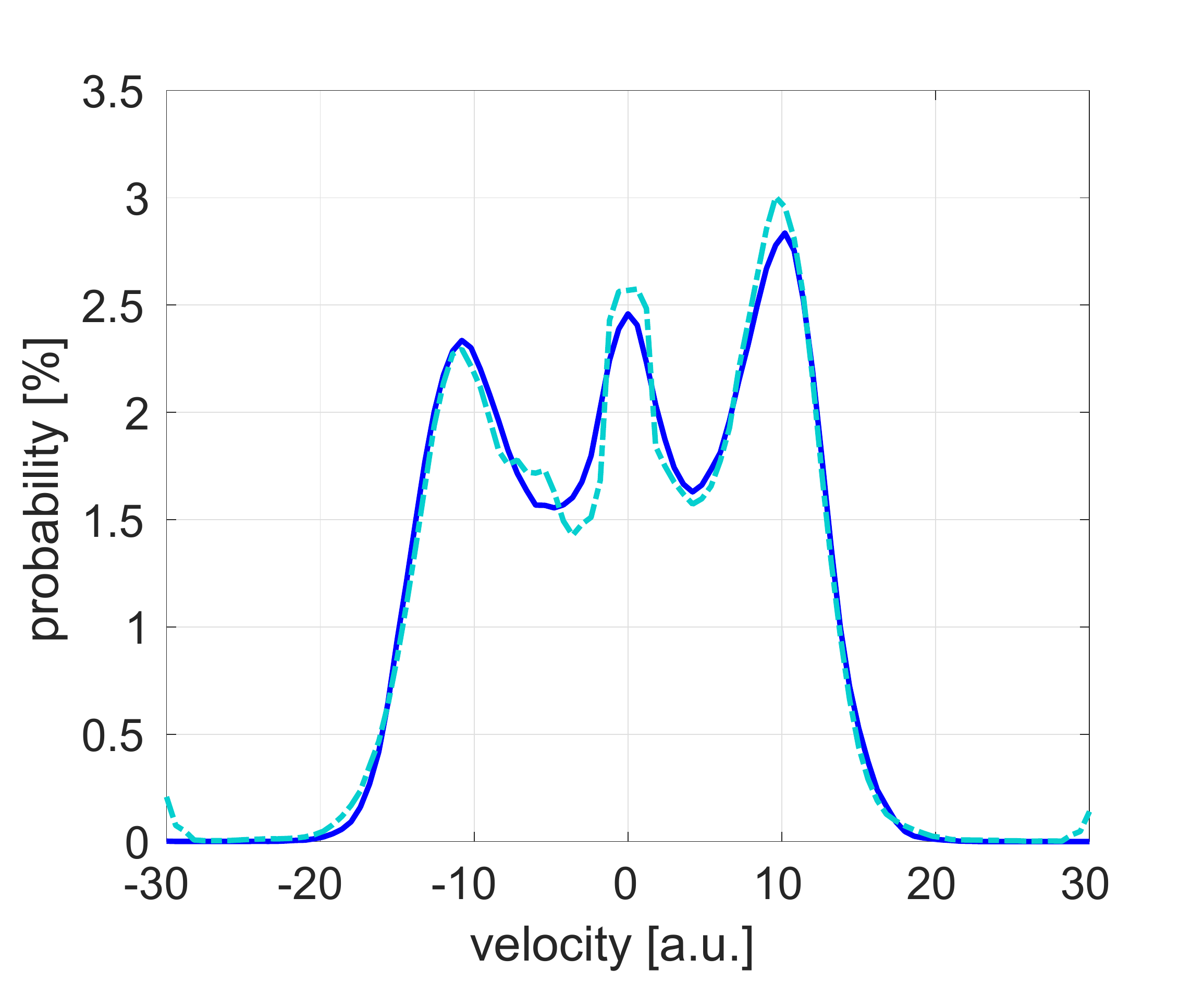}
		    \includegraphics[width=0.44\textwidth, trim={0 0 0 35},clip]{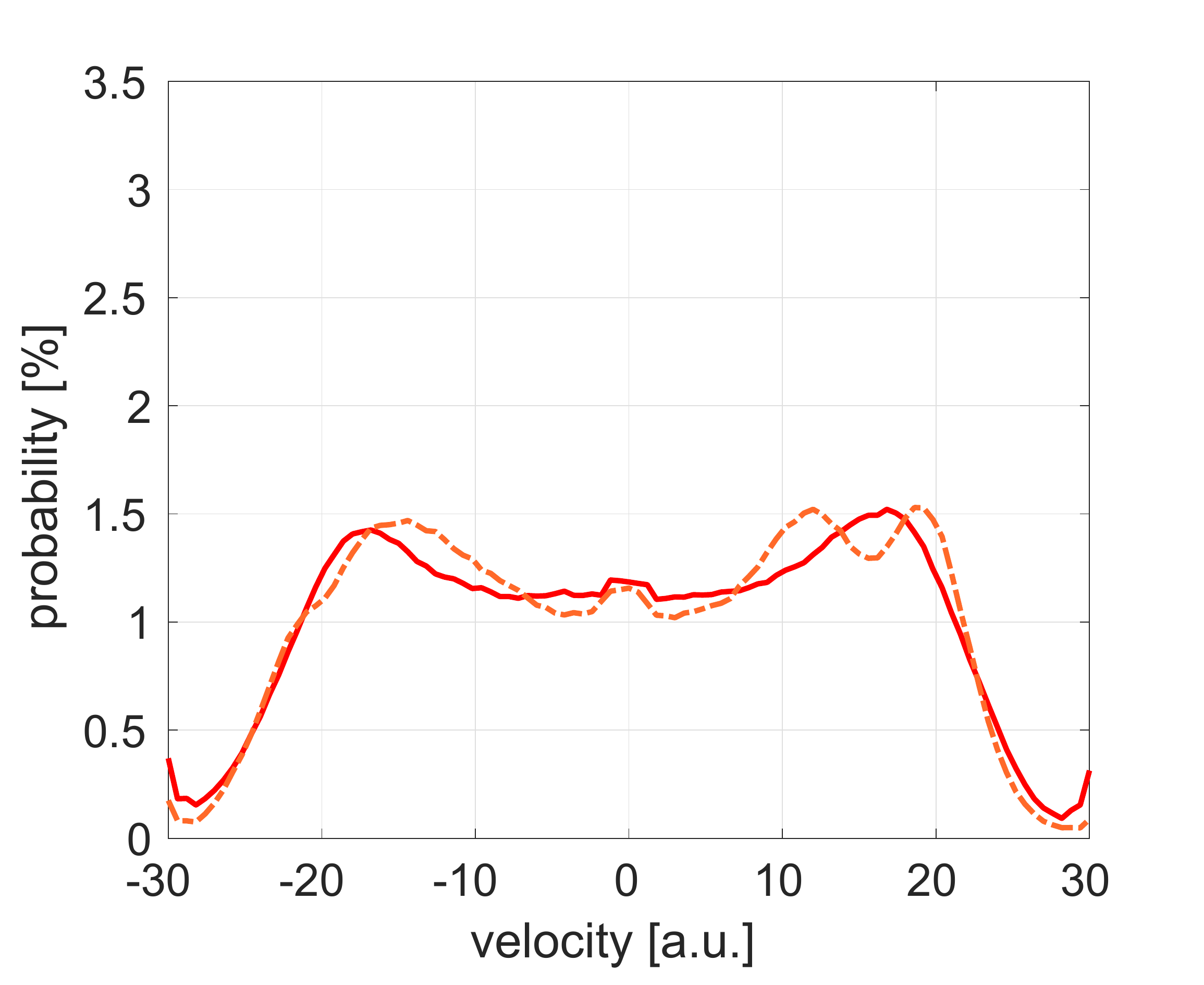}\\
		    \includegraphics[width=0.44\textwidth]{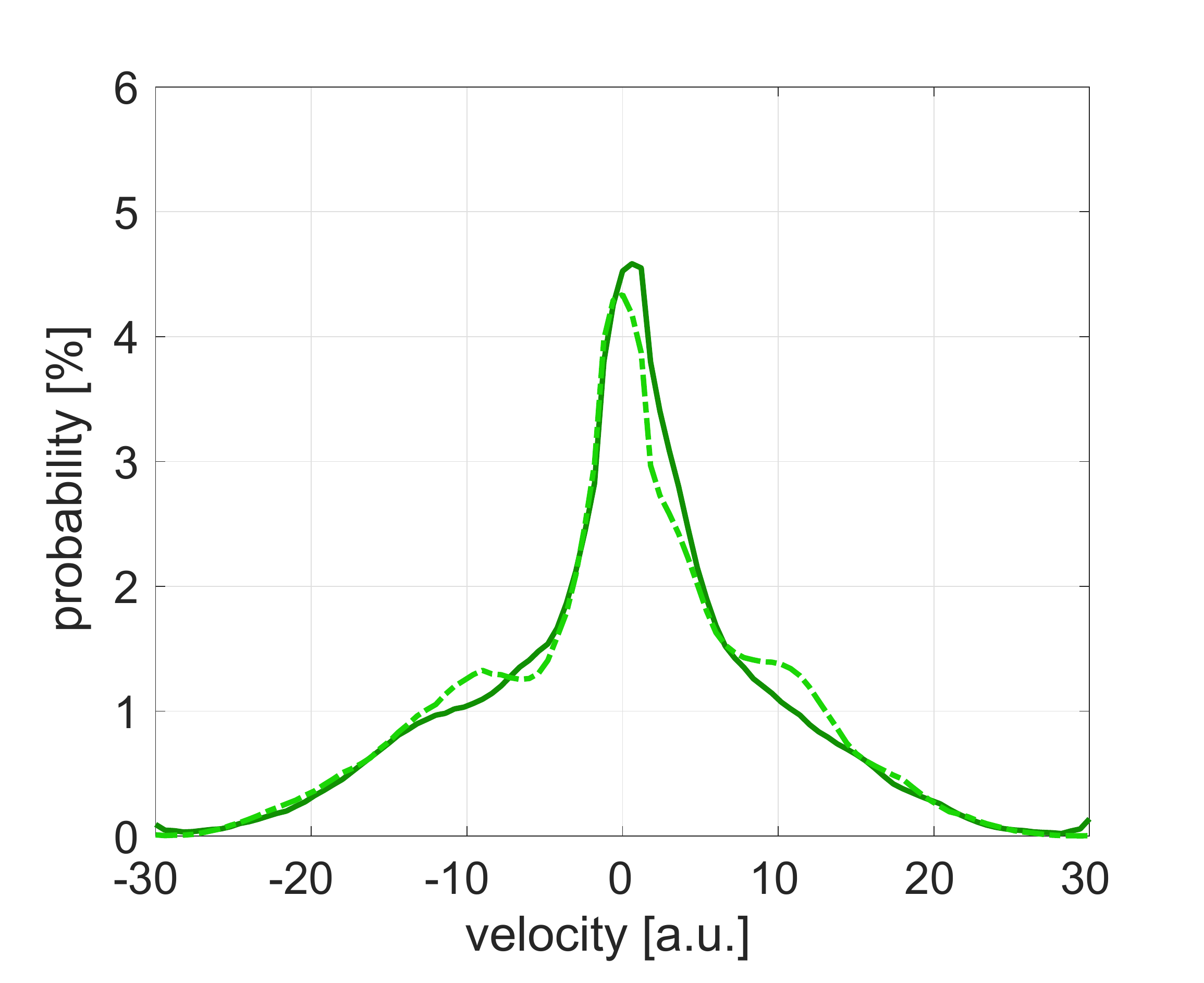}}}
		\caption{IMS of three different human players (in blue, red and green) are plotted together with the IMS produced by their corresponding MC-based IMS Generator (dash-dotted light blue, orange and light green lines).}
		\label{fig:pdf_MC}
\end{figure*}

For the sake of brevity, Fig. \ref{fig:pdf_MC} shows the representative comparison between the synthetic IMS and those of computed from the HP motion for 3 out of 6 IMS Generators that were synthesised (the other three showing similar properties).

\begin{figure*}[thpb]
    \framebox{\parbox{0.97\textwidth}{
			\centering
			\includegraphics[width=0.45\textwidth]{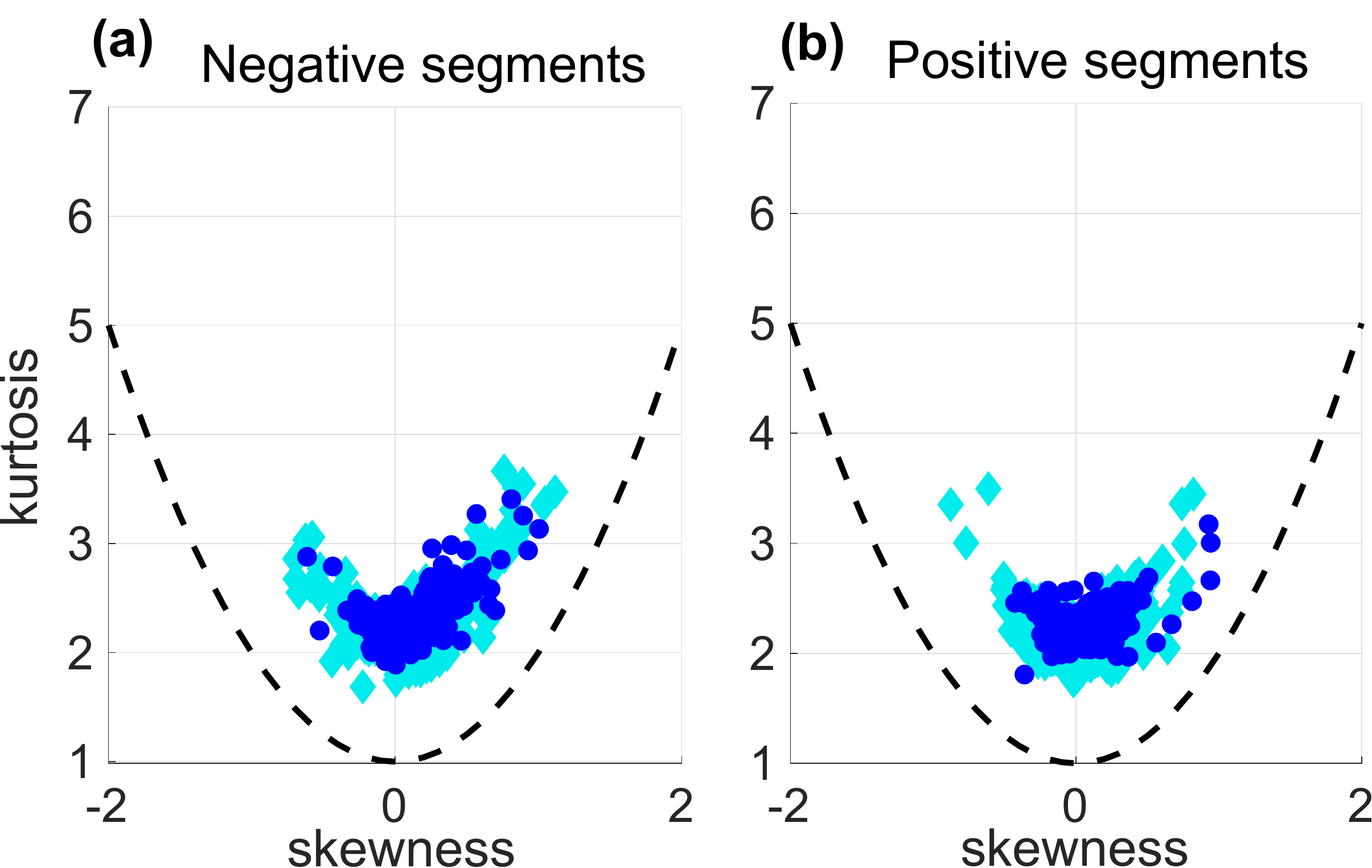}
			\hspace{10pt}
	        \includegraphics[ width=0.45\textwidth]{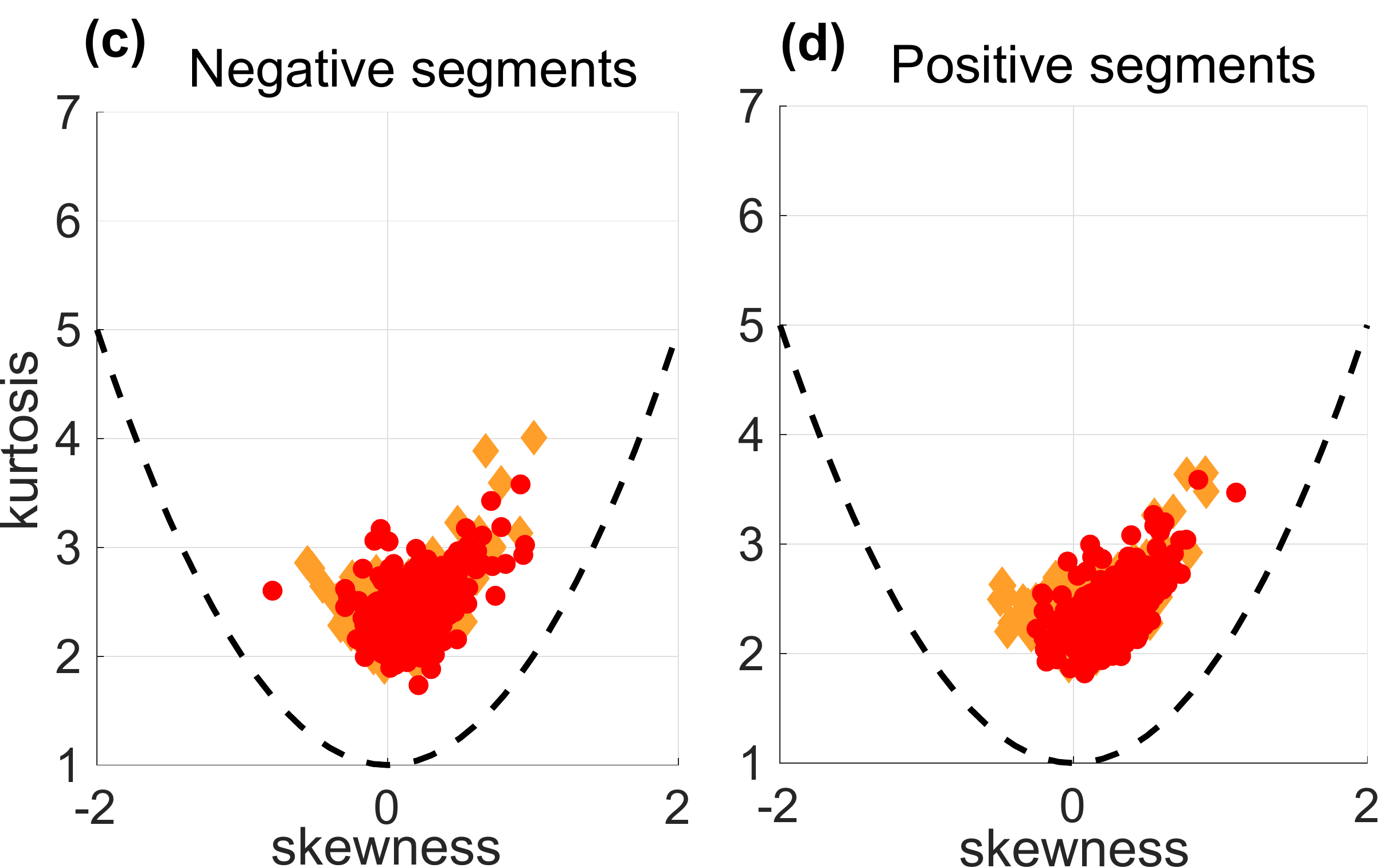}\\
	        \vspace{10pt}
	        \includegraphics[width=0.45\textwidth]{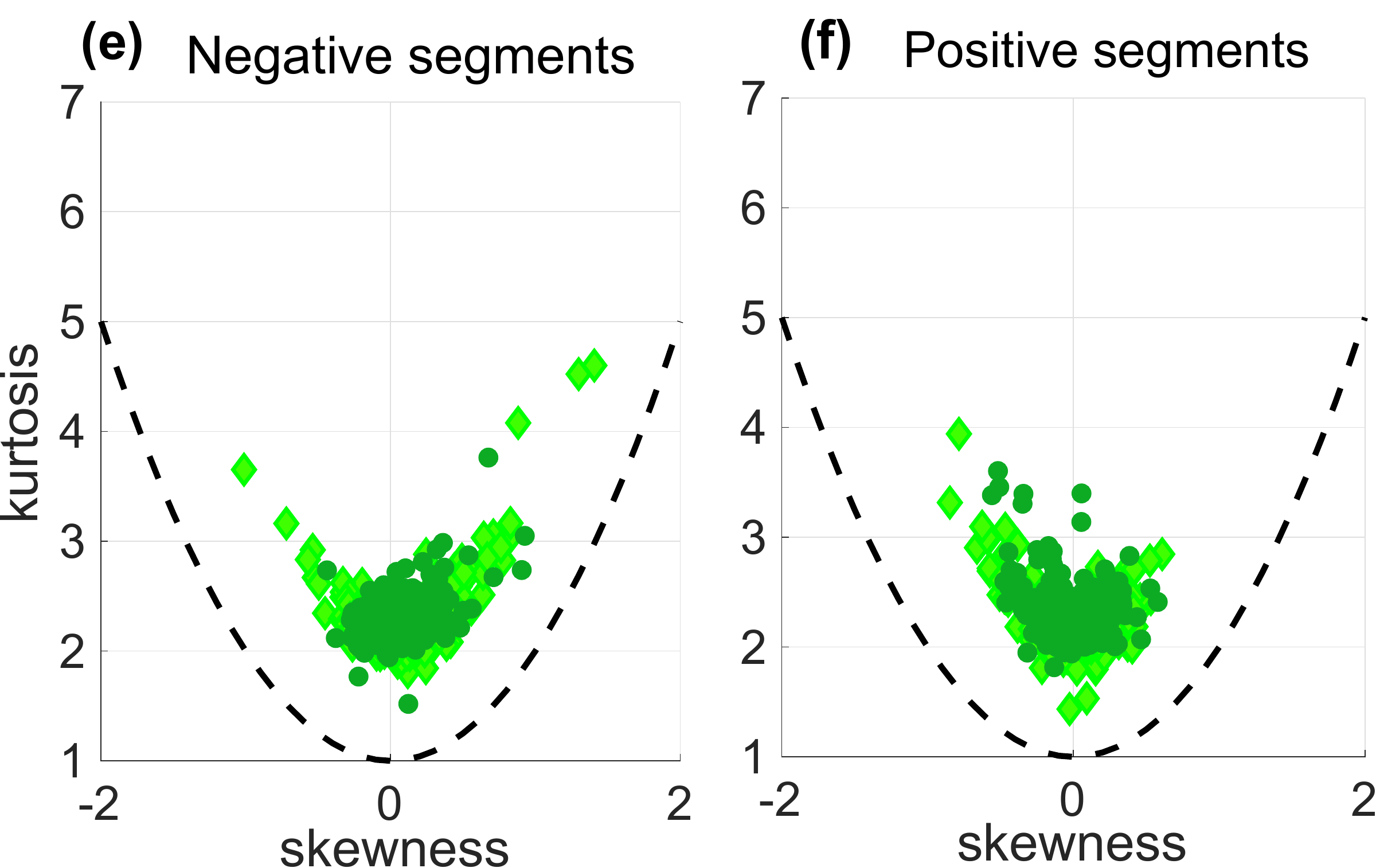}}}
	\caption{Analysis of the trajectories generated by human players and the IMS generators in the skewness-kurtosis plane.For each human player two different plots are reported for negative \textbf{(a), (c), (e)} and positive \textbf{(b), (d), (f)} segments. The velocity segments belonging to the three human players are represented respectively as points in dark blue, red and green whereas the generated ones as diamonds in light blue, orange and light green.}
	\label{fig:ks_plane_MC}
\end{figure*}

\begin{figure*}[thpb]
	\framebox{\parbox{0.97\textwidth}{
			\centering
			\vspace{8pt}
			\includegraphics[width=0.88\textwidth]{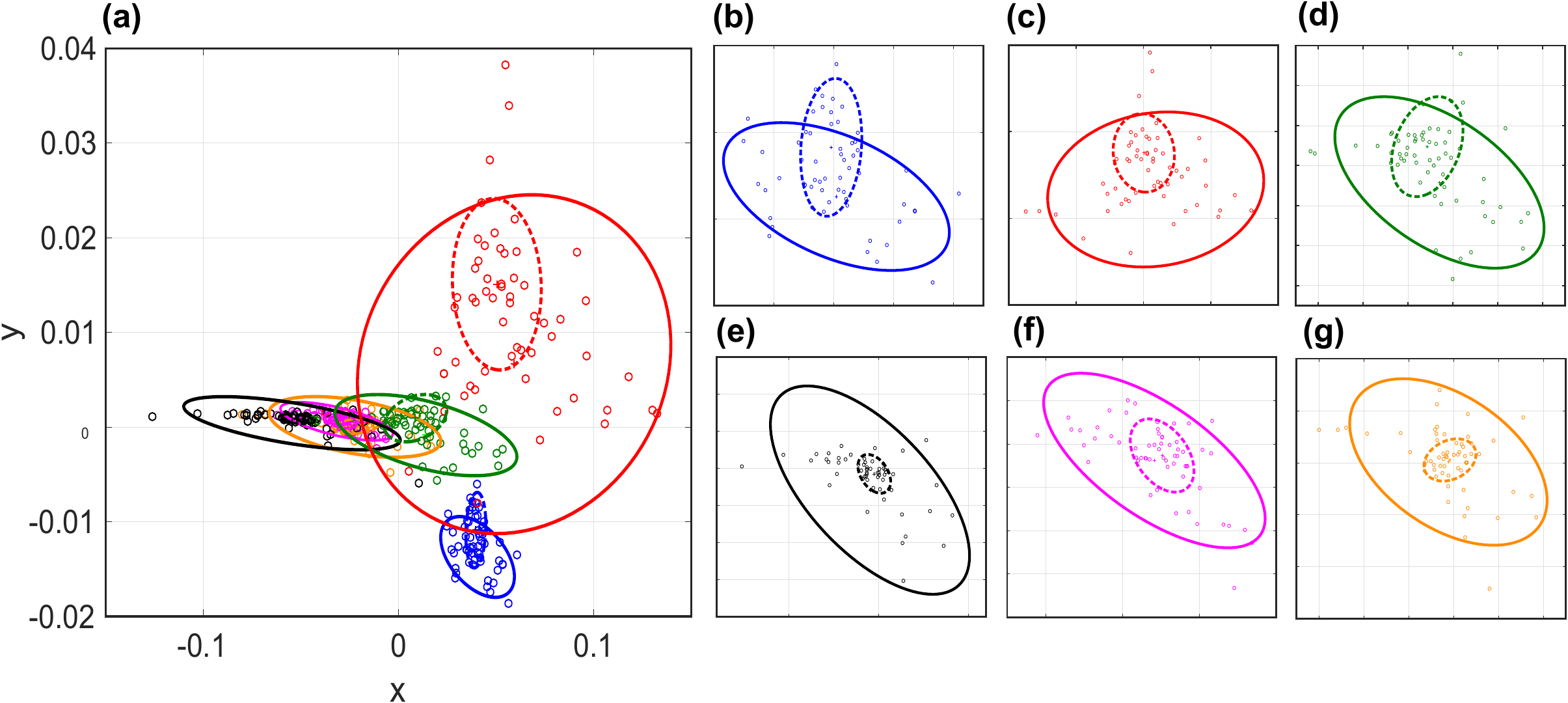}}}
	\vspace{8pt}
	\caption{Similarity space. Each trial is represented as a point in the plane, an ellipse with the same colour encircles all the points belonging to the same agent. A different colour is used for each couple human player - Markov chain. \textbf{(a)} All characteristic regions belonging to the human players (continuous line) are represented together with the regions defined by the MCs (dashed line). \textbf{(b)-(g)} Six different plots show the similarity space for each player (continuous line) individually with the corresponding artificial signatures (dashed line).}
	\label{fig:similarity_space_MC}
\end{figure*}

To evaluate the skewness and the kurtosis of the generated signals, the velocity time series were divided into segments; each corresponding to one direction of motion (left-to-right or right-to-left). Each segment was rescaled to a common support ($\left[0,1\right]$) and normalised so that its underneath area is unitary. Velocity segments with less than $20$ samples were ignored. 
The skewness and kurtosis of each segment is represented as a point in Fig. \ref{fig:ks_plane_MC} both for the data obtained by the HPs participating in the experiments and for the synthetic data generated by the MCs trained on the same HPs. We observe that the velocity segments generated by the MCs are located in the same area of the skewness-kurtosis plane occupied by the human players.

Fig. \ref{fig:similarity_space_MC} shows the mapping of the IMS Generator outputs and the corresponding human players' motion in the similarity space. The overlapping between the ellipses corresponding to human player trials with those generated by the Markov chain confirms the effectiveness of the modelling approach and the ability of the IMS generators to generate human-like synthetic movement signals that can be used to remove the need of any prerecorded signature in the cognitive architecture schematic presented in Fig. \ref{fig:control_architecture}.

\subsection{Validation of the Virtual Trainers}
To evaluate the performance of the virtual trainers based on the architecture presented in Sec. \ref{sec:vp_modelling}, we tested their ability to track the motion of another player while exhibiting the desired IMS encoded by the IMS Generator it incorporated. 

The VT was validated when playing the mirror game both as a follower and a leader  in different dyadic sessions against a human player. As a representative case we show here the case where the IMS Generator used in the cognitive architecture driving the VT is that derived from time series acquired from player $2$ playing in solo condition which was validated earlier in Sec. \ref{sec:modelling}.
The parameters of the optimal control strategy in \ref{eq:controller} are tuned heuristically as in \cite{Alderisio2016,Zhai2017}. Specifically: $\alpha = 1, \beta = 2, \gamma = -1, \eta = 10^{-4}$ and with $T = 0.03s$. For the VT to act as a leader, $\theta_p$ is set to $0.1$ and $\omega = 0.8$, whereas for it to act as a follower, $\theta_p = 0.9$ and $\omega = 0.1$.

We then evaluate the performance and accuracy of the VT by computing the following metrics.
\begin{enumerate}
	\item The \textit{Root mean square error} (RMSE) is used to evaluate the tracking error between the position time series of the two players given by
	\begin{equation}
	RMSE = \sqrt{\frac{1}{N} \sum_{k=1}^{N} \left({r_p}_{k} - x_{k}\right)^2},
	\end{equation}
	where $N$ is the number of samples in the simulation, ${r_p}_{k}$ and $x_{k} $ are the position values at the $k$th time instant of the HP and the VT respectively.
	\item The \textit{Relative phase} (RP) defined as the difference between the phases of the players: $\Delta\Phi := \Phi_{HP} - \Phi_{VT}$ is used as an  indicator to distinguish leading from following behaviour, as commonly done in the movement science literature \cite{Alderisio2018}. The phase was computed from the data by using an Hilbert transform as proposed in \cite{Kralemann2008}.  A positive value of $\Delta\Phi$ indicates that the HP is moving ahead of the VT, implying that the HP is leading while the VT is following. Viceversa, negative values of  $\Delta\Phi$ imply that the HP is following the VT.
	\item \textit{Circular variance} (CV) is used to quantify the coordination level between the two players and is computed as \cite{Kreuz2007}
	\begin{equation}
	CV = \left\|\frac{1}{N} \sum_{k=1}^{N}e^{i\Delta\Phi_k}  \right\| \quad \in \left[0,1\right],
	\end{equation}
	where $N$ is the number of collected samples, $\Delta\Phi_k$ is the relative phase between the two players at a time instant $k$ and $\left\| \cdot \right\|$ denote the 2-norm. The higher the CV, the higher the coordination level among the players is.
\end{enumerate}

\begin{figure*}[tb]
	\framebox{\parbox{0.99\textwidth}{
			\centering
			\vspace{8pt}
			\includegraphics[width=0.94\textwidth]{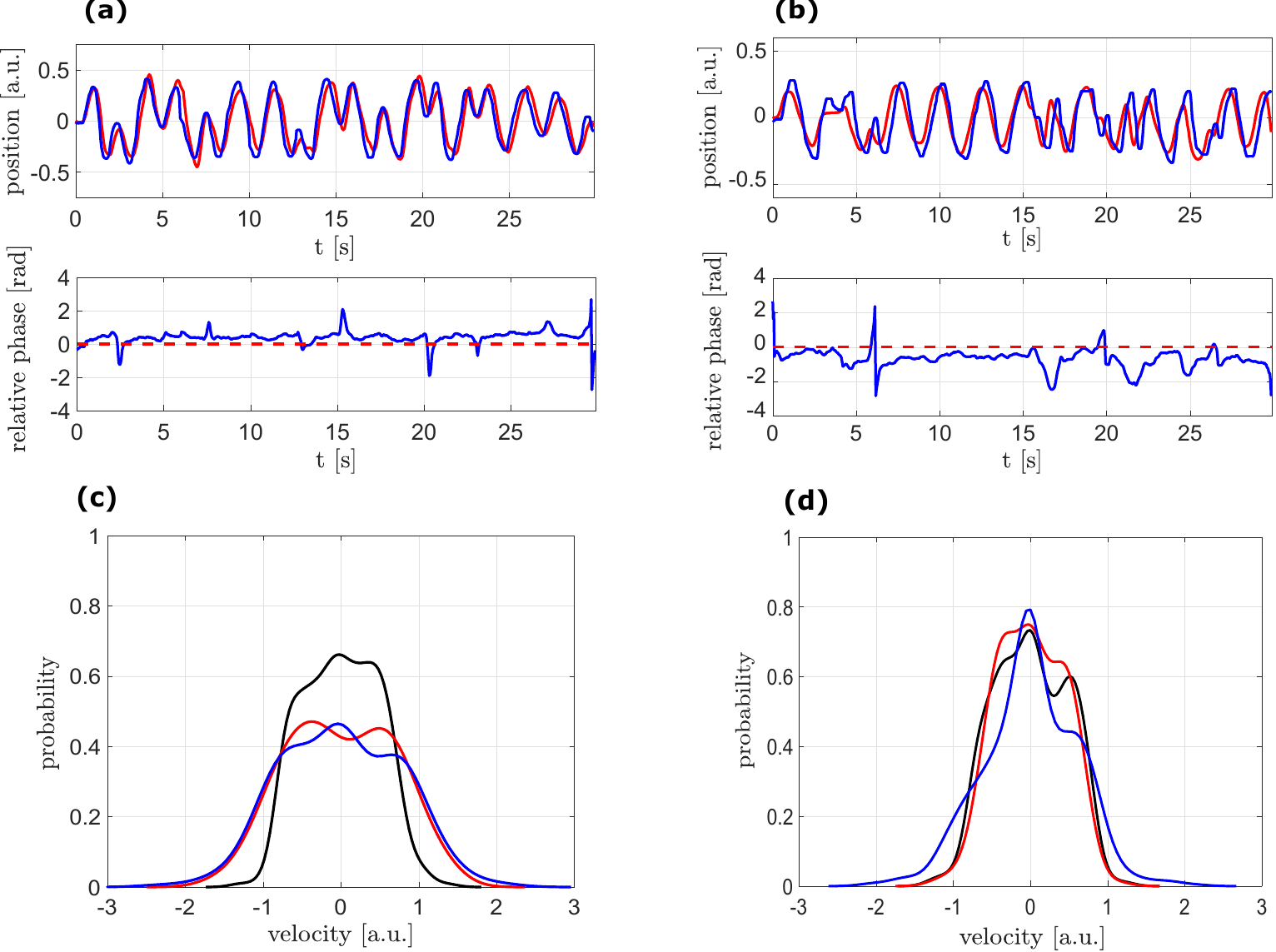}}}
	\vspace{8pt}
	\caption{Session of mirror game in leader-follower condition. Position time series, relative phase \textbf{(a)} and velocity distribution \textbf{(c)} between the two players with VT acting as follower. Position time series, relative phase \textbf{(b)} and velocity distribution \textbf{(d)} between the two players with VT acting as leader. In red the HP, in blue the VT and in black the IMS generated by the MC and given as reference.}
	\label{fig:validation_VP}
\end{figure*}

Fig. \ref{fig:validation_VP} shows the results of two representative experiments where the VT is playing as a follower [panels (a) and (c)] or a leader [panels (b) and (d)]. In the former case,  the CV between the two players was computed to be $0.933$ which indicates the VT reaches a high level of coordination with the HP. This is confirmed by the RMS of the position error which is found to be only $0.112$. The positive relative phase between the two players in panel (a) confirms that the VT is indeed acting as a follower while the HP as a leader ($\Delta \Phi = 0.394 \pm 0.408$).  When the VT acts as a leader, we find again good coordination levels with the CV reaching $0.868$ and the RMS of the position error being $0.122$. In this case the relative phase shift is negative, indicating that the VT is effectively acting as a leader as desired ($\Delta \Phi = -0.664 \pm 0.574$).

As depicted in Fig. \ref{fig:validation_VP}(c) and Fig. \ref{fig:validation_VP}(d), the VT adapts its IMS to reach a compromise between that of player 2 in solo condition and that of the player it is playing against. When the VT plays as a leader [panel (d)] we see that the mismatch between the reference IMS and that of the VT becomes smaller forcing the HP to adapt its IMS to that of the VT. This is confirmed by our computations of the EMD between the velocity profiles of player 2 (used as a reference) and those of the VT and the HP. Namely we find $EMD\left(Ref,VT\right)= 0.006; EMD(HP, VT) = 0.018$ when the VT plays as leader, whereas when it plays as a follower  $EMD\left(Ref,VT\right)= 0.0263; EMD(HP, VT) = 0.011$.

\subsection{Validation of the RL-based Cyberplayer}
Next, we validate the approach presented in Sec. \ref{sec:cp_modelling}. As mentioned therein, synthetic data is used for training which is obtained by running the mirror game between two virtual trainers. 

Specifically, a set of 6 VTs (VT$_1$, VT$_2$, VT$_3$, VT$_4$, VT$_5$, VT$_6$) were synthesised, each possessing an IMS Generator parameterized after each of the $6$ human participants (using sessions of the mirror game in the solo condition as described in Sec. \ref{sec:modelling}).

As a testbed example, we then set the goal of training the CP to act as a follower (or a leader) during the game while behaving as HP$_5$. To this aim we made VT$_1$ to VT$_4$ to play the mirror game as leaders (or as followers) against VT$_5$ so as to generate enough synthetic data for the training. 

\begin{figure*}[thpb]
	\framebox{\parbox{0.98\textwidth}{
			\centering
			\vspace{8pt}
			\includegraphics[width=0.94\textwidth]{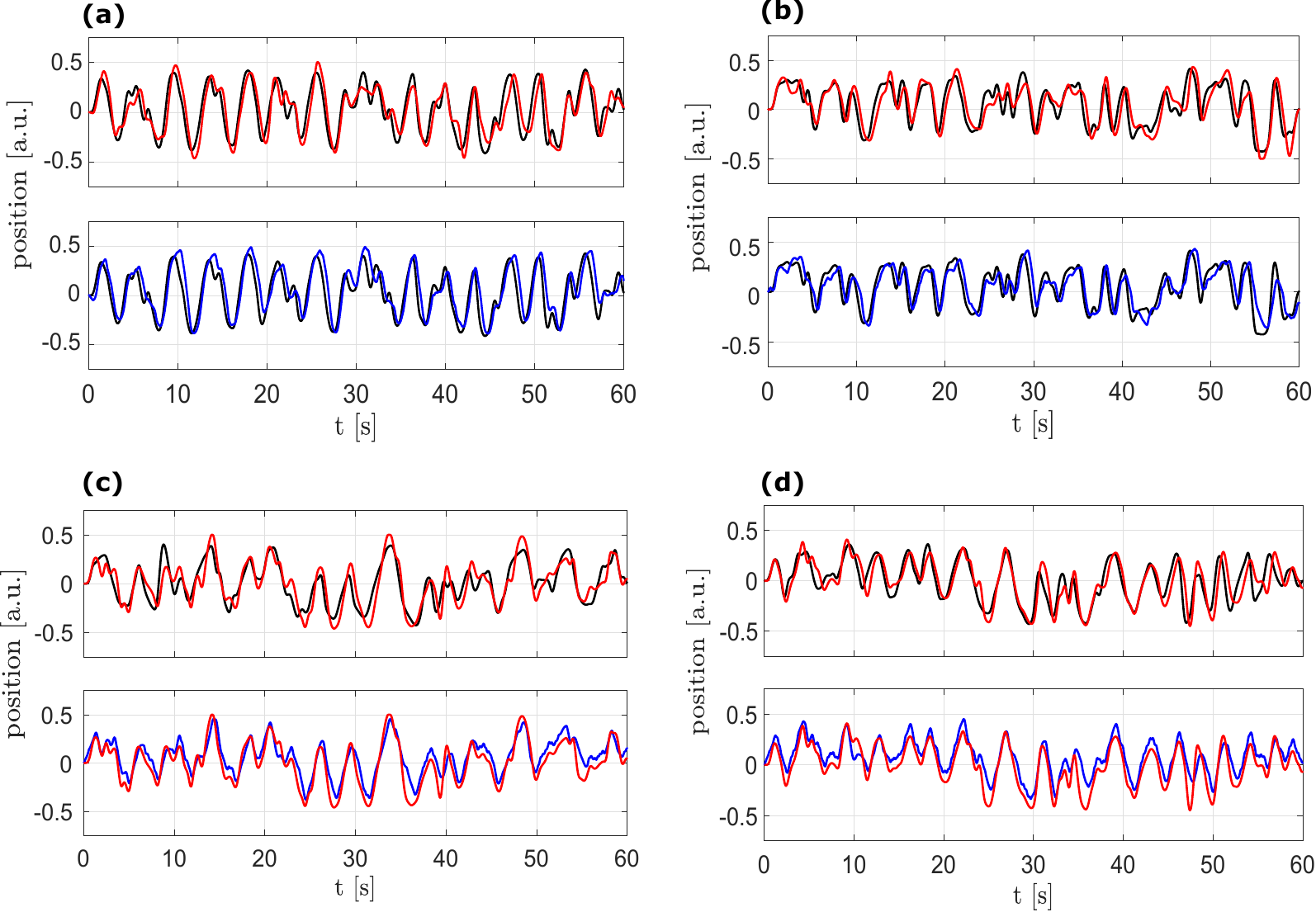}}}
	\vspace{8pt}
	\caption{Position time series of a dyadic session of the CP trained as follower against a leader both \textbf{(a)} included and \textbf{(b)} not included in the training set. Position time series of a dyadic session of the CP trained as leader against a follower both \textbf{(c)} included and \textbf{(d)} not included in the training set. The CP is in blue, the VT follower in red and the VT leader in black.}
	\label{fig:validation_CP}
\end{figure*}

\begin{figure}[thpb]
	\framebox{\parbox{0.97\columnwidth}{
			\centering
			\vspace{8pt}
			\includegraphics[width=0.9\columnwidth]{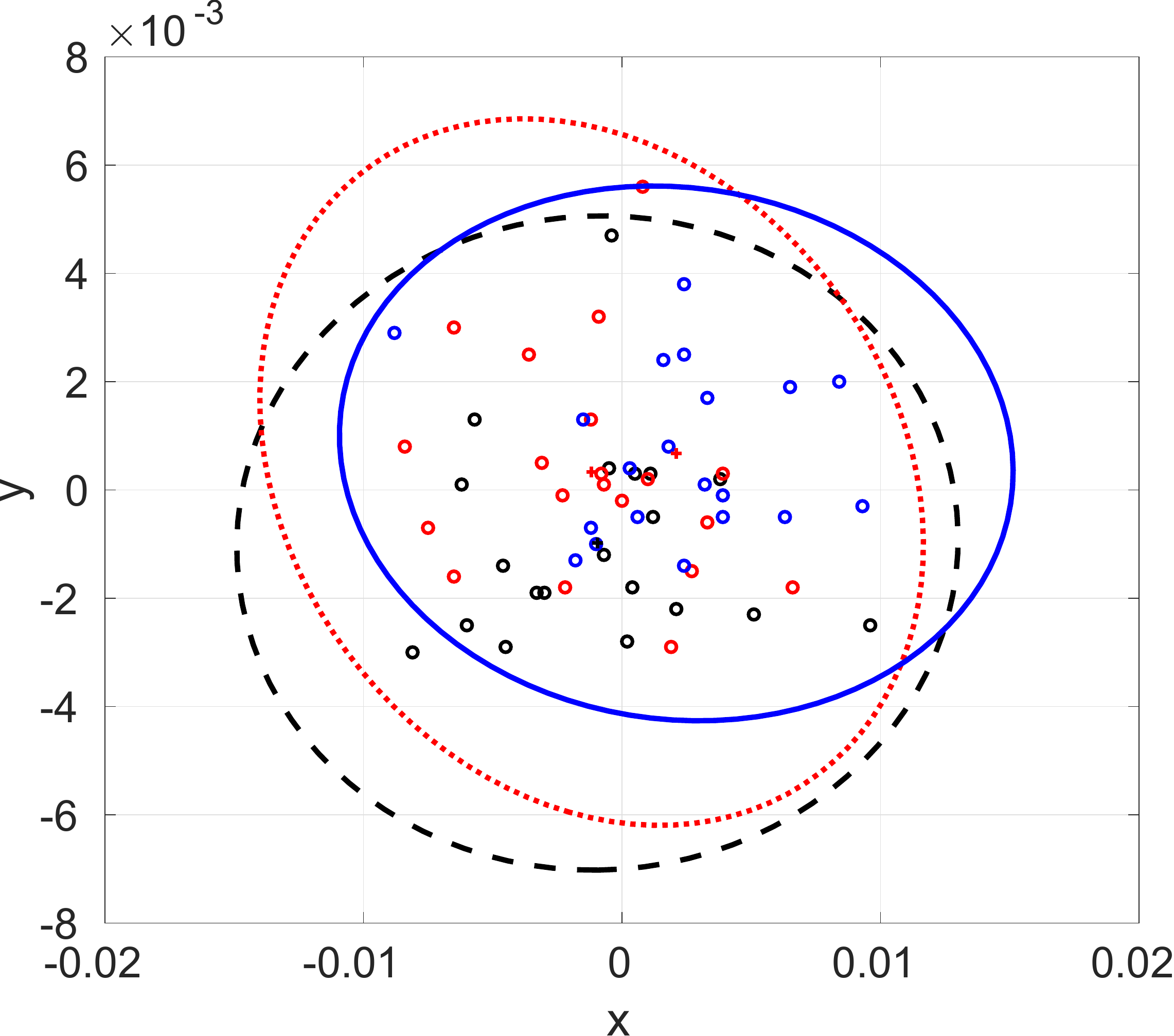}}}
	\caption{Similarity Space. Characteristic regions built with $20$ sessions of mirror game in leader-follower configuration. The same leader (in black) plays first with the virtual trainer (in red) and then with the cyber player (in blue).}
	\label{fig:similaritySpace_CP}
\end{figure}

The CP was  trained by observing each pair of VTs playing against each other for approximately $5$ days (playing approximately $24000$ trials of the mirror game).

The performance was then evaluated by comparing the trained CP behaviour with that of the target VT while playing with the VTs it was trained upon and with another VT not included in the training phase ($VT_6$). Specifically, validation was carried out over $20$ sessions of the game with and without the CP. A representative dataset is shown in Fig. \ref{fig:validation_CP} where each panel shows sessions of the mirror game played by two VTs versus sessions of the game where the CP replaces one of them (as a follower, panels (a) and (b), or a leader, panels (c) and (d)). 
We clearly see that the CP is able to effectively replace the model-based VT both in the case where it plays against a VT included in the training set (panels (a) and (c)) that against a new one (panels (b) and (d)).

To check whether the IMS exhibited by the CP during the game matches that of the players it has learnt from, we show in  Fig. \ref{fig:similaritySpace_CP} the characteristic regions of the following VT (in red) and of the CP (in blue) obtained by acquiring data during $20$ sessions of the game each plays against the same virtual trainer acting as a leader (whose characteristic region is depicted in black). The figure shows a high degree of overlap between the IMS of the CP and that of the VT it was trained upon confirming the effectiveness of this approach.

Mathematically, we computed the overlap $A_{ij}$ between the two ellipses, say ellipse $i$ and ellipse $j$, as the ratio between the area of their intersection and their total area. In this way, a complete overlap between the two ellipses corresponds to $A_{ij} = 1$, whilst $A_{ij} = 0$ corresponds to complete disjoint sets. We obtained overlap values equal or greater than $0.6$ for all the pairs of players we analysed ($A_{VT_L,VT_F} = 0.74, \; A_{CP,VT_L}=0.60,\; A_{CP,VT_F}=0.62$) confirming that they share similar kinematic features.
Quantitatively, we also computed the EMDs between different pairs of players including the human players that the IMS generators used in the model-based VTs were modelled upon. We found that the $EMD\left(HP, VT\right) = 0.0024$, $EMD\left(HP, CP\right) = 0.0033$,  and a $RMSE\left(HP, VT\right) = 0.104 \pm 0.006$ and $RMSE\left(HP, CP\right) = 0.1088 \pm 0.015$ which confirms the tracking ability of the CP when compared to that of the finely tuned  model-based VT it used to train itself upon.

\begin{table}[thpb]
    \renewcommand{\arraystretch}{1.2}
    \caption{Earth mover's distance (EMD), circular variance (CV) and root mean square of the position error (RMS) are reported for the CP/VT$_5$ while playing with each of the other players. Two different scenarios are considered: CP as follower and as leader.}
    \label{tab:cp_vp_experiments}
    \centering
    \subtable{
         \begin{tabular}{|c|c|c|c|c|}
             \cline{2-5}
             \multicolumn{1}{c|}{} & \textbf{Following} & \textbf{EMD} & \textbf{CV} & \textbf{RMS} \\
             \hline
             \multirow{2}{*}{VT$_1$} & VT$_5$ & $0.002$ & $0.76 \pm 0.06$ & $0.11 \pm 0.01$ \\
             \cline{2-5}
             & CP & $0.008$ & $0.80 \pm 0.05$ &  $0.11 \pm 0.01$ \\
             \hhline{|=|=|=|=|=|}
             \multirow{2}{*}{VT$_2$} & VT$_5$ & $0.003$ & $0.86 \pm 0.06$ & $0.11 \pm 0.01$ \\
             \cline{2-5}
             & CP & $0.004$ & $0.86 \pm 0.06$ &  $0.12 \pm 0.01$ \\
             \hhline{|=|=|=|=|=|}
             \multirow{2}{*}{VT$_3$} & VT$_5$ & $0.003$ & $0.85 \pm 0.04$ & $0.12 \pm 0.01$ \\
             \cline{2-5}
             & CP & $0.007$ & $0.88 \pm 0.04$ &  $0.12 \pm 0.01$ \\
             \hhline{|=|=|=|=|=|}
             \multirow{2}{*}{VT$_4$} & VT$_5$ & $0.004$ & $0.74 \pm 0.07$ & $0.10 \pm 0.01$ \\
             \cline{2-5}
             & CP & $0.004$ & $0.79 \pm 0.07$ &  $0.10 \pm 0.01$ \\
             \hhline{|=|=|=|=|=|}
             \multirow{2}{*}{VT$_6$} & VT$_5$ & $0.003$ & $0.83 \pm 0.05$ & $0.11 \pm 0.01$ \\
             \cline{2-5}
             & CP & $0.002$ & $0.86 \pm 0.04$ &  $0.11 \pm 0.01$ \\
             \hline
         \end{tabular}
    }
    \vspace{16pt}
    \subtable{
        \begin{tabular}{|c|c|c|c|c|}
             \cline{2-5}
             \multicolumn{1}{c|}{} & \textbf{Leading} & \textbf{EMD} & \textbf{CV} & \textbf{RMS} \\
             \hline
             \multirow{2}{*}{VT$_1$} & VT$_5$ & $0.003$ & $0.83 \pm 0.04$ & $0.11 \pm 0.01$ \\
             \cline{2-5}
             & CP & $0.007$ & $0.88 \pm 0.03$ &  $0.12 \pm 0.01$ \\
             \hhline{|=|=|=|=|=|}
             \multirow{2}{*}{VT$_2$} & VT$_5$ & $0.003$ & $0.82 \pm 0.03$ & $0.12 \pm 0.01$ \\
             \cline{2-5}
             & CP & $0.007$ & $0.90 \pm 0.03$ &  $0.11 \pm 0.01$ \\
             \hhline{|=|=|=|=|=|}
             \multirow{2}{*}{VT$_3$} & VT$_5$ & $0.003$ & $0.81 \pm 0.05$ & $0.12 \pm 0.01$ \\
             \cline{2-5}
             & CP & $0.007$ & $0.90 \pm 0.03$ &  $0.12 \pm 0.01$ \\
             \hhline{|=|=|=|=|=|}
             \multirow{2}{*}{VT$_4$} & VT$_5$ & $0.003$ & $0.86 \pm 0.05$ & $0.10 \pm 0.01$ \\
             \cline{2-5}
             & CP & $0.007$ & $0.89 \pm 0.03$ &  $0.11 \pm 0.01$ \\
             \hhline{|=|=|=|=|=|}
             \multirow{2}{*}{VT$_6$} & VT$_5$ & $0.003$ & $0.84 \pm 0.04$ & $0.10 \pm 0.01$ \\
             \cline{2-5}
             & CP & $0.007$ & $0.90 \pm 0.03$ &  $0.11 \pm 0.01$ \\
             \hline
        \end{tabular}
    }
\end{table}

\begin{figure*}[thpb]
	\framebox{\parbox{0.98\textwidth}{
			\centering
			\vspace{8pt}
			\includegraphics[width=0.96\textwidth]{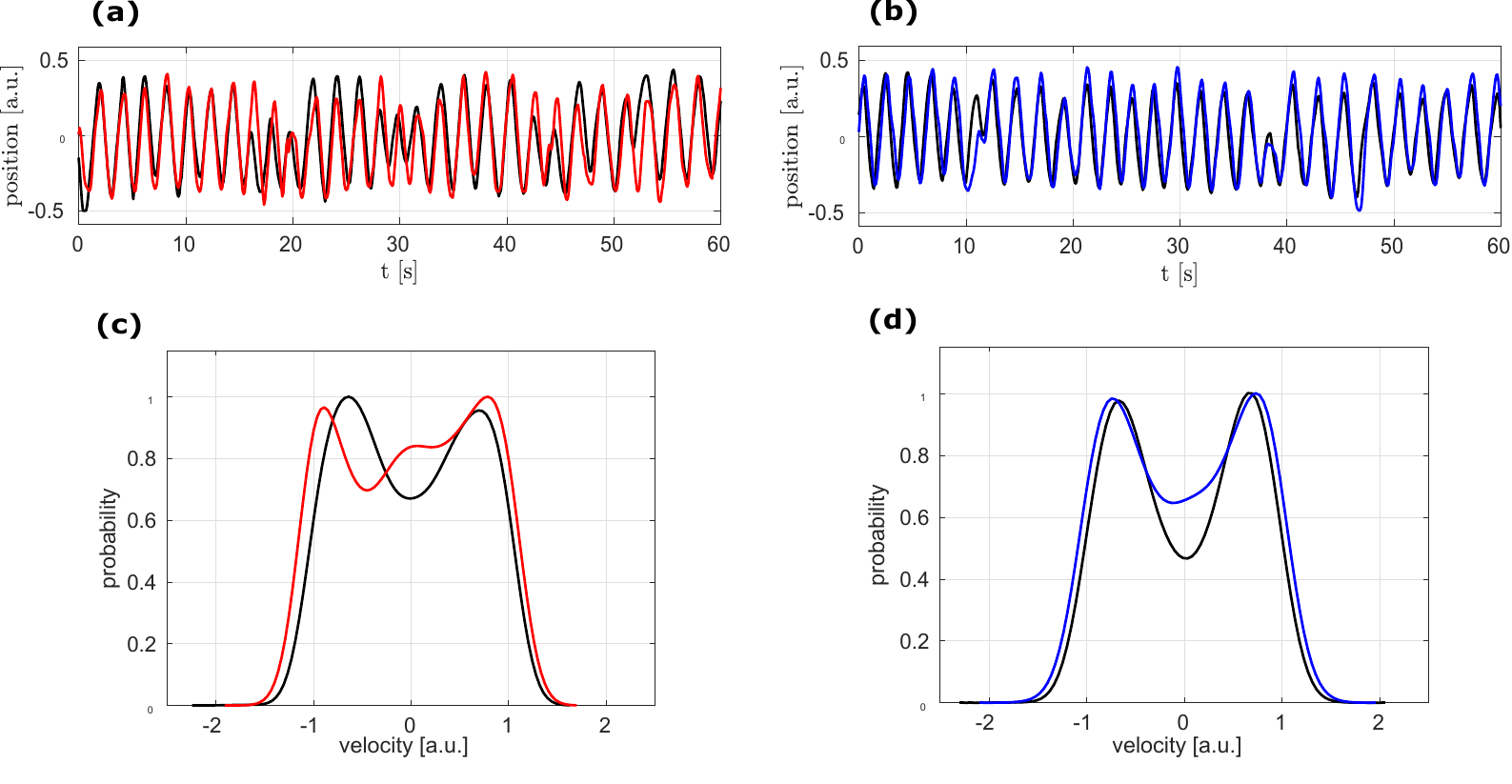}}}
	\vspace{8pt}
	\caption{Position time series \textbf{(a)} and velocity PDF \textbf{(c)} of a leader-follower trial between the human leader (in black) and the human follower (in red). Position time series \textbf{(b)} and velocity PDF \textbf{(d)} of a leader-follower trial between the human leader (in black) and the cyber player follower (in blue).}
	\label{fig:HP_HP_complete}
\end{figure*}

Further quantitative evidence is provided in Table \ref{tab:cp_vp_experiments} where EMD, CV and RMS of the position error between the CP and the VTs modelled on players $1,2,3,4$ and $6$ are given and respectively compared to those obtained when the same VTs are made to play against the VT modelled after player $5$ which the CP was trained to imitate. The lack of any statistically significant difference between the CP and the VT it was trained upon confirms that the training phase has been successful. 

To collect further evidence that the CP trained using synthetic data generated by the VT modelled after human player 5 is able to autonomously play the game as them, we performed an additional sets of experiments where HP$_5$ and the CP are each made to play the mirror game with 4 different HPs (players from $1$ to $4$) acting as leaders.  A total of $8$ trials were performed for each pair and the results compared against each other. Each trial was $60$ seconds long and was performed through CHRONOS \cite{Alderisio2017b}. In Fig. \ref{fig:HP_HP_complete} the representative position time series and IMS of a  mirror game session where a human leader and a human follower play against each other is compared with the time series and IMS acquired during a session where  the CP is made to play the game as a follower against the same human leader.
Quantitatively, for each pair of players we evaluated the distance between the corresponding PDFs by means of EMD. Other indices considered in the comparison are related to the level of synchronisation measured in terms of CV and RMS. Results are reported in Table \ref{tab:cp_experiments}, where the mean and the standard deviation are reported over the total number of trials for each metric and for each pair of players. Furthermore paired T-test was performed at $95\%$ confidence level with $N-1$ degrees of freedom. As reported Table \ref{tab:t-test}, a p-value $> 0.05$ was computed for each pair of players showing  that there is no significant difference between the behaviour of the cyber player when playing the game and that of the human player whose Markov Chain model was used by the VT during the training stage. This implies that the CP has learnt successfully to adapt its IMS and playing behaviour to that of the HP as desired.

\begin{table}[thpb]
    \renewcommand{\arraystretch}{1.2}
    \caption{Earth mover's distance (EMD), circular variance (CV) and root mean square of the position error (RMS) are reported for each pair of players. The human players are numbered from $1$ to $5$, player $5$ is the human follower while players $1-4$ are human leaders.}
    \label{tab:cp_experiments}
    \centering
    \begin{tabular}{|c|c|c|c|c|}
         \cline{2-5}
         \multicolumn{1}{c|}{} & \textbf{Following} & \textbf{EMD} & \textbf{CV} & \textbf{RMS} \\
         \hline
         \multirow{2}{*}{HP$_1$} & HP$_5$ & $0.01$ & $0.9 \pm 0.05$ & $0.18 \pm 0.03$ \\
         \cline{2-5}
         & CP & $0.01$ & $0.9 \pm 0.06$ &  $0.14 \pm 0.06$ \\
         \hhline{|=|=|=|=|=|}
         \multirow{2}{*}{HP$_2$} & HP$_5$ & $0.007$ & $0.84 \pm 0.07$ & $0.15 \pm 0.04$ \\
         \cline{2-5}
         & CP & $0.006$ & $0.89 \pm 0.05$ &  $0.15 \pm 0.05$ \\
         \hhline{|=|=|=|=|=|}
         \multirow{2}{*}{HP$_3$} & HP$_5$ & $0.006$ & $0.92 \pm 0.01$ & $0.17 \pm 0.03$ \\
         \cline{2-5}
         & CP & $0.005$ & $0.93 \pm 0.06$ &  $0.15 \pm 0.05$ \\
         \hhline{|=|=|=|=|=|}
         \multirow{2}{*}{HP$_4$} & HP$_5$ & $0.009$ & $0.86 \pm 0.06$ & $0.16 \pm 0.03$ \\
         \cline{2-5}
         & CP & $0.004$ & $0.87 \pm 0.03$ &  $0.12 \pm 0.02$ \\
         \hline
    \end{tabular}
\end{table}

\begin{table}[thpb]
    \renewcommand{\arraystretch}{1.2}
    \caption{Paired T-test between all pairs human leader/human follower and human leader/cyber player has been performed both for the CV and RMS.}
    \label{tab:t-test}
    \centering
    \begin{tabular}{|p{14mm}|p{19mm}|p{19mm}|}
         \hline
         \qquad \textbf{Pair} & \qquad \textbf{CV} & \qquad \textbf{RMS} \\
         \hline
         HP$_1$ -- HP$_5$ \qquad HP$_1$ -- CP & $t(7)=-0.081$, $p=0.938$ & $ t(7)=1.562$, $p=0.162$\\
         \hline
         HP$_2$ -- HP$_5$ \qquad HP$_2$ -- CP & $t(7)=-2.105$, $p=0.073$ & $t(7)=-0.347$, $p=0.739$ \\
         \hline
         HP$_3$ -- HP$_5$ \qquad HP$_3$ -- CP & $t(7)=-0.496$, $p=0.635$ & $t(7)=1.25$, $p=0.251$ \\
         \hline
         HP$_4$ -- HP$_5$ \qquad HP$_4$ -- CP & $t(7)=0.933$, $p=0.382$ & $t(7)=2.340$, $p=0.052$ \\
         \hline
    \end{tabular}
\end{table}

\section{Conclusion}
\label{sec:conclusions}
We addressed the problem of designing control architectures to construct artificial agents able to engage with humans in motor coordination tasks. The mirror game was chosen as a paradigmatic example where two individuals have to perform a joint oscillatory task. To overcome the limitations of previous approaches, we proposed the use of Markov chains to model human behaviour in the game and remove the need of pre-recorded human trajectories in previous control approaches in the literature. We showed that Markov chain models can be obtained that reproduce the unique kinematic features (IMS) that distinguish the motion of different people. We then embedded such Markov chain models in a control architecture that achieves the goal of making a virtual agent able to coordinate its motion with that of a human player in a completely autonomous manner.

Still the problem remained of having to fine tune the parameters of the deterministic  control algorithm used to drive the artificial agent based on solving an optimal control problem on a receding horizon. To overcome this problem, we proposed a  cognitive architecture based on reinforcement learning able to make a cyberplayer  play the mirror game with another player in different conditions. 

To train such a Cyberplayer one would need in theory a relatively large amount of data from sessions of the game played by human players against each other. When this is not available we propose that model-based artificial agents (or Virtual Trainers) endowing the IMS generator derived in this paper can be used during the training to generate as much synthetic data as needed for the RL algorithm to converge.

After presenting the results of the training, we validated the cyber player both numerically and experimentally and showed that it is indeed able to coordinate its motion with that of another (human or virtual) player while exhibiting the IMS of a target individual. Moreover, the RL-based cyberplayer, once trained, was shown to be able to play with other players that were not used for its training without requiring the lengthy offline parameter tuning needed to synthesise artificial model-based mirror game players.

We wish to emphasise that the proposed cyber player can be a promising tool to be used in the human dynamic clamp setting proposed in \cite{Dumas2014} as a method to study social interaction and movement coordination among humans.
Also, the CP we developed in this paper can be effectively used to implement the rehabilitation strategies for patients affected by social disorders that were suggested in \cite{Bardy2014,Slowinski2017}. Therein, it is suggested that for rehabilitation purposes the patient should be made to play the mirror game with an artificial avatar that starts by playing sessions of the game while exhibiting the same IMS of the patient and is then gradually made to change its IMS to habituate the patient to coordinate its motion with players moving in a different way from her/him. Our approach allows to easily realise such an avatar. Indeed, the patient IMS could be captured by our model-based Markov Chain approach. The resulting IMS generator can then be used to synthesise VTs able to train the CP to play the game as the patient during the initial sessions. Further VTs with dissimilar IMS from those of the patient can then be used to re-train the CP for later sessions of the rehabilitation exercise implementing the vision first proposed in \cite{Bardy2014,Slowinski2016,Slowinski2017} where the IMS of the avatar is changed from that of the patient during the first mirror game sessions to become that of a healthy individual towards the end of the rehabilitation \cite{Bardy2014,Dumas2014}. In this way, the patient is habituated over successive trials of the game with the CP to coordinate his/her motion better and better before moving to  trials involving other humans. 
We wish to emphasise that such clinical application of the CP is beyond the scope of this paper, which is focused on the development of the IMS generator, the virtual trainers and the RL based CP. The potential application of the CP in a clinical setting will be the subject of future work.

The approach we presented was effective at solving the problem of achieving motor coordination between the CP and the HP in a single DOF motor task. Nevertheless, the proposed AI-based architecture is general enough to be easily extended to a multi DOF problem where the state space and the reward function will depend upon the position $\mathbf{x}$ of the player end-effector as a point $[x,y,z]$ in Cartesian coordinates. The main drawback for these problems is that the state space to be explored by the learning algorithm may become quickly too large to be dealt with by tabular methods such as the Q-learning algorithm. A possible solution is the use of different learning strategies such as deep reinforcement learning \cite{Mnih2015}, which can be more apt to tackle the challenging case of higher DOF tasks. A preliminary investigation of the application of deep reinforcement learning to drive avatars in human motor coordination task was recently discussed in \cite{Lombardi2019} and is the subject of ongoing research.

\ifCLASSOPTIONcaptionsoff
  \newpage
\fi

\addtolength{\textheight}{-6cm}   
\vspace{0.5cm}

\bibliography{library.bib}

\begin{thebibliography}{10}
\providecommand{\url}[1]{#1}
\csname url@samestyle\endcsname
\providecommand{\newblock}{\relax}
\providecommand{\bibinfo}[2]{#2}
\providecommand{\BIBentrySTDinterwordspacing}{\spaceskip=0pt\relax}
\providecommand{\BIBentryALTinterwordstretchfactor}{4}
\providecommand{\BIBentryALTinterwordspacing}{\spaceskip=\fontdimen2\font plus
\BIBentryALTinterwordstretchfactor\fontdimen3\font minus
  \fontdimen4\font\relax}
\providecommand{\BIBforeignlanguage}[2]{{%
\expandafter\ifx\csname l@#1\endcsname\relax
\typeout{** WARNING: IEEEtran.bst: No hyphenation pattern has been}%
\typeout{** loaded for the language `#1'. Using the pattern for}%
\typeout{** the default language instead.}%
\else
\language=\csname l@#1\endcsname
\fi
#2}}
\providecommand{\BIBdecl}{\relax}
\BIBdecl

\bibitem{Iqbal2016}
T.~Iqbal, S.~Rack, and L.~D. Riek, ``{Movement Coordination in Human-Robot
  Teams: A Dynamical Systems Approach},'' \emph{IEEE Transactions on Robotics},
  vol.~32, no.~4, pp. 909--919, 2016.

\bibitem{Atkeson2000}
C.~G. Atkeson, J.~G. Hale, F.~Pollick, M.~Riley, S.~Kotosaka, S.~Schaal,
  T.~Shibata, G.~Tevatia, A.~Ude, S.~Vijayakumar, E.~Kawato, and M.~Kawato,
  ``{Using humanoid robots to study human behavior},'' \emph{IEEE Intelligent
  Systems and their Applications}, vol.~15, no.~4, pp. 46--56, 2000.

\bibitem{Shukla2012}
A.~Shukla and A.~Billard, ``{Coupled dynamical system based arm-hand grasping
  model for learning fast adaptation strategies},'' \emph{Robotics and
  Autonomous Systems}, vol.~60, no.~3, pp. 424--440, 2012.

\bibitem{Richardson2007}
M.~J. Richardson, K.~L. Marsh, R.~W. Isenhower, J.~R. Goodman, and R.~C.
  Schmidt, ``{Rocking together: Dynamics of intentional and unintentional
  interpersonal coordination},'' \emph{Human Movement Science}, vol.~26, no.~6,
  pp. 867--891, 2007.

\bibitem{Neda2000}
Z.~N{\'{e}}da, E.~Ravasz, Y.~Brechet, T.~Vicsek, and A.~L. Barab{\'{a}}si,
  ``{The sound of many hands clapping},'' \emph{Nature}, vol. 403, no. 6772,
  pp. 849--850, 2000.

\bibitem{Wing1995}
A.~M. Wing and C.~Woodburn, ``{The coordination and consistency of rowers in a
  racing eight},'' \emph{Journal of Sports Sciences}, vol.~13, no.~3, pp.
  187--197, 1995.

\bibitem{Codrons2014}
E.~Codrons, N.~F. Bernardi, M.~Vandoni, and L.~Bernardi, ``{Spontaneous group
  synchronization of movements and respiratory rhythms},'' \emph{PLoS ONE},
  vol.~9, no.~9, p. e107538, 2014.

\bibitem{Noy2011}
L.~Noy, E.~Dekel, and U.~Alon, ``{The mirror game as a paradigm for studying
  the dynamics of two people improvising motion together},'' \emph{Proceedings
  of the National Academy of Sciences}, vol. 108, no.~52, pp. 20\,947--20\,952,
  2011.

\bibitem{Slowinski2016}
P.~S{\l}owi{\'{n}}ski, C.~Zhai, F.~Alderisio, R.~Salesse, M.~Gueugnon,
  L.~Marin, B.~G. Bardy, M.~di~Bernardo, and K.~Tsaneva-Atanasova, ``{Dynamic
  similarity promotes interpersonal coordination in joint action},''
  \emph{Journal of The Royal Society Interface}, vol.~13, no. 116, p. 20151093,
  2016.

\bibitem{Slowinski2017}
P.~S{\l}owi{\'{n}}ski, F.~Alderisio, C.~Zhai, Y.~Shen, P.~Tino, C.~Bortolon,
  D.~Capdevielle, L.~Cohen, M.~Khoramshahi, A.~Billard, R.~Salesse,
  M.~Gueugnon, L.~Marin, B.~G. Bardy, M.~di~Bernardo, S.~Raffard, and
  K.~Tsaneva-Atanasova, ``{Unravelling socio-motor biomarkers in
  schizophrenia},'' \emph{npj Schizophrenia}, vol.~3, no.~1, p.~8, 2017.

\bibitem{DelMonte2013}
J.~Del-Monte, D.~Capdevielle, M.~Varlet, L.~Marin, R.~Schmidt, R.~Salesse
  \emph{et~al.}, ``Social motor coordination in unaffected relatives of
  schizophrenia patients: a potential intermediate phenotype,'' \emph{Frontiers
  in Behavioural Neuroscience}, vol.~7, no. 137, 2013.

\bibitem{Varlet2012}
M.~Varlet, L.~Marin, S.~Raffard, R.~C. Schmidt, D.~Capdevielle, J.~P.
  Boulenger, J.~Del-Monte, and B.~G. Bardy, ``{Impairments of social motor
  coordination in schizophrenia},'' \emph{PLoS ONE}, vol.~7, no.~1, 2012.

\bibitem{Raffard2015}
S.~Raffard, R.~N. Salesse, L.~Marin, J.~Del-Monte, R.~C. Schmidt, M.~Varlet,
  B.~G. Bardy, J.-P. Boulenger, and D.~Capdevielle, ``Social priming enhances
  interpersonal synchronization and feeling of connectedness towards
  schizophrenia patients,'' \emph{Scientific reports}, vol.~5, p. 8156, 2015.

\bibitem{Wainer2014}
J.~Wainer, B.~Robins, F.~Amirabdollahian, and K.~Dautenhahn, ``Using the
  humanoid robot kaspar to autonomously play triadic games and facilitate
  collaborative play among children with autism,'' \emph{IEEE Transactions on
  Autonomous Mental Development}, vol.~6, no.~3, pp. 183--199, 2014.

\bibitem{Begum2016}
M.~Begum, R.~W. Serna, and H.~A. Yanco, ``Are robots ready to deliver autism
  interventions? a comprehensive review,'' \emph{International Journal of
  Social Robotics}, vol.~8, no.~2, pp. 157--181, 2016.

\bibitem{Zhai2015}
C.~Zhai, F.~Alderisio, K.~Tsaneva-Atanasova, and M.~di~Bernardo, ``A model
  predictive approach to control the motion of a virtual player in the mirror
  game,'' in \emph{54th IEEE Annual Conference on Decision and Control (CDC)},
  2015, pp. 3175--3180.

\bibitem{Zhai2016}
C.~Zhai, F.~Alderisio, P.~Slowinski, K.~Tsaneva-Atanasova, and M.~di~Bernardo,
  ``{Design of a Virtual Player for Joint Improvisation with Humans in the
  Mirror Game},'' \emph{PLoS ONE}, vol.~11, no.~4, p. e0154361, 2016.

\bibitem{Zhai2017}
------, ``{Design and Validation of a Virtual Player for Studying Interpersonal
  Coordination in the Mirror Game},'' \emph{IEEE Transactions on Cybernetics},
  vol.~48, pp. 1018--1029, 2017.

\bibitem{Haken1985}
H.~Haken, J.~A.~S. Kelso, and H.~Bunz, ``{A Theoretical Model of Phase
  Transitions in Human Hand Movements},'' \emph{Biol. Cybern}, vol.~51, pp.
  347--356, 1985.

\bibitem{Alderisio2017b}
F.~Alderisio, M.~Lombardi, G.~Fiore, and M.~di~Bernardo, ``{A novel
  computer-based set-up to study movement coordination in human ensembles},''
  \emph{Frontiers in Psychology}, vol.~8, p. 967, 2017.

\bibitem{Chronos}
\BIBentryALTinterwordspacing
M.~Lombardi. Chronos: a tool to study synchronization and coordination in human
  ensembles. [Online]. Available:
  \url{https://dibernardogroup.github.io/Chronos/index.html}
\BIBentrySTDinterwordspacing

\bibitem{Russel2003}
S.~Russel and P.~Norvig, \emph{{Artificial Intelligence: A Modern Approach}},
  3rd~ed.\hskip 1em plus 0.5em minus 0.4em\relax Prentice Hall, 2003.

\bibitem{Sutton1998}
R.~S. Sutton, A.~G. Barto \emph{et~al.}, \emph{Reinforcement learning: An
  introduction}.\hskip 1em plus 0.5em minus 0.4em\relax MIT press, 1998.

\bibitem{White1989}
C.~C. White~III and D.~J. White, ``Markov decision processes,'' \emph{European
  Journal of Operational Research}, vol.~39, no.~1, pp. 1--16, 1989.

\bibitem{Rabiner1989}
L.~Rabiner, ``{A tutorial on hidden Markov models and selected applications in
  speech recognition},'' \emph{Proceedings of the IEEE}, vol.~77, no.~2, pp.
  257--286, 1989.

\bibitem{Ahmed2012}
N.~Ahmed and K.~R. Rao, \emph{{Orthogonal transforms for digital signal
  processing}}.\hskip 1em plus 0.5em minus 0.4em\relax Springer Science \&
  Business Media, 2012.

\bibitem{Heinzel2002}
G.~Heinzel, A.~R{\"{u}}diger, R.~Schilling, and T.~Hannover, ``{Spectrum and
  spectral density estimation by the Discrete Fourier transform (DFT),
  including a comprehensive list of window functions and some new flat-top},''
  Tech. Rep., 2002.

\bibitem{Rabiner1975}
L.~R. Rabiner and B.~Gold, \emph{Theory and application of digital signal
  processing}.\hskip 1em plus 0.5em minus 0.4em\relax Englewood Cliffs, NJ,
  Prentice-Hall, Inc., 1975.

\bibitem{Leapmotion}
\BIBentryALTinterwordspacing
Ultraleap. Leap motion controller. [Online]. Available:
  \url{https://www.ultraleap.com/product/leap-motion-controller/}
\BIBentrySTDinterwordspacing

\bibitem{Borg2012}
I.~Borg, P.~J. Groenen, and P.~Mair, \emph{Applied multidimensional
  scaling}.\hskip 1em plus 0.5em minus 0.4em\relax Springer Science \& Business
  Media, 2012.

\bibitem{Alderisio2016}
F.~Alderisio, D.~Antonacci, C.~Zhai, and M.~Bernardo, ``{Comparing Different
  Control Approaches to Implement a Human-like Virtual Player in the Mirror
  Game},'' in \emph{European Control Conference (ECC)}, vol. 216.\hskip 1em
  plus 0.5em minus 0.4em\relax IEEE, 2016, pp. 216--221.

\bibitem{Alderisio2018}
F.~Alderisio, M.~Lombardi, and M.~di~Bernardo, ``{Emergence of leadership in
  complex networks and human groups},'' in \emph{2018 IEEE International
  Symposium on Circuits and System (ISCAS)}, 2018, pp. 1--5.

\bibitem{Kralemann2008}
B.~Kralemann, L.~Cimponeriu, M.~Rosenblum, A.~Pikovsky, and R.~Mrowka, ``{Phase
  dynamics of coupled oscillators reconstructed from data},'' \emph{Physical
  Review E - Statistical, Nonlinear, and Soft Matter Physics}, vol.~77, no.~6,
  pp. 1--16, 2008.

\bibitem{Kreuz2007}
T.~Kreuz, F.~Mormann, R.~G. Andrzejak, A.~Kraskov, K.~Lehnertz, and
  P.~Grassberger, ``{Measuring synchronization in coupled model systems: A
  comparison of different approaches},'' \emph{Physica D: Nonlinear Phenomena},
  vol. 225, no.~1, pp. 29--42, 2007.

\bibitem{Dumas2014}
G.~Dumas, G.~C. de~Guzman, E.~Tognoli, and J.~A.~S. Kelso, ``{The human dynamic
  clamp as a paradigm for social interaction},'' \emph{Proceedings of the
  National Academy of Sciences}, vol. 111, no.~35, pp. E3726--E3734, 2014.

\bibitem{Bardy2014}
B.~G. Bardy, R.~N. Salesse, M.~Gueugnon, Z.~Zhong, J.~Lagarde, and L.~Marin,
  ``Movement similarities and differences during social interaction: The
  scientific foundation of the alterego european project,'' in \emph{2014 IEEE
  International Conference on Systems, Man, and Cybernetics (SMC)}.\hskip 1em
  plus 0.5em minus 0.4em\relax IEEE, 2014, pp. 772--777.

\bibitem{Mnih2015}
V.~Mnih, K.~Kavukcuoglu, D.~Silver, A.~Rusu, J.~Veness, M.~Bellemare,
  A.~Graves, M.~Riedmiller, and A.~Fidjeland, ``{Human-level control through
  deep reinforcement learning},'' \emph{Nature Letter}, vol. 518, pp. 529--533,
  2015.

\bibitem{Lombardi2019}
M.~Lombardi, D.~Liuzza, and M.~di~Bernardo, ``Deep learning control of
  artificial avatars in group coordination tasks,'' in \emph{2019 IEEE
  International Conference on Systems, Man and Cybernetics (SMC)}, 2019, pp.
  724 -- 729.

\end{thebibliography}

\begin{IEEEbiography}[{\includegraphics[width=1in,height=1.25in,clip,keepaspectratio]{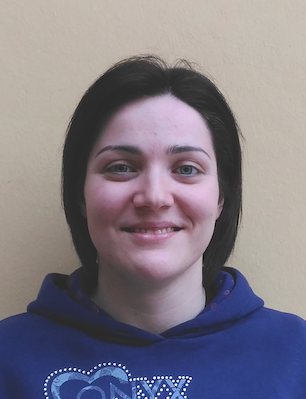}}]{Maria Lombardi} (M'18) received her M.Sc. degree in Computer Engineering in 2016 from the University of Naples ``Federico" (Italy). She is currently a Ph.D student in Engineering Mathematics at University of Bristol (UK).
	
	Her current research interests include human-robot interaction, design of artificial agents interacting with humans as well as modelling coordination and leadership both in human groups and in groups mixed with virtual agents.
	
	She is a member of Alumni and Friends association at University of Bristol and of the IEEE Robotics and Automation Society.
\end{IEEEbiography}

\begin{IEEEbiography}[{\includegraphics[width=1in,height=1.25in,clip,keepaspectratio]{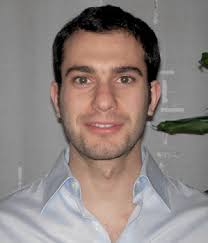}}]{Davide Liuzza} received the PhD. degree in Automation Engineering in 2013 from the University of Naples ``Federico II" (Italy).
	
	He was a visiting Ph.D student at the Department of Applied Mathematics at the University of Bristol (UK) in 2012, and at the ACCESS Linnaeus Centre, Royal Institute of Technology (KTH), Stockholm (Sweden) from 2012 to 2013. From 2013 to 2015 he was a post-doctoral researcher at the Automatic Control Laboratory at KTH. From 2016 to 2018 he was a post-doctoral researcher with the department of Engineering at University of Sannio in Italy. In 2018 he was a visiting researcher at the Mechatronics Research Group at Chalmers University of Technology, Gothenburg, Sweden. 
	
	He is currently a researcher at the Fusion and Nuclear Safety Department at ENEA (Italy), where he got a permanent research position.

    His research interests are in networked control systems, coordination of multi-agent systems, incremental stability of nonlinear systems, as well as nonlinear control and hybrid systems. He is also interested in energy and nuclear fusion systems control.
\end{IEEEbiography}

\begin{IEEEbiography}[{\includegraphics[width=1in,height=1.25in,clip,keepaspectratio]{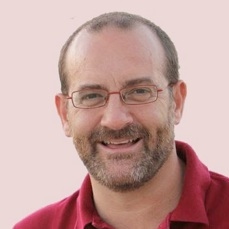}}]{Mario di Bernardo} (SM'06--F'12) is Professor of Automatic Control at the University of Naples ``Federico II", Italy. He is also Honorary Professor at Fudan University, Shanghai in China and Professor (part-time) of Nonlinear Systems and Control at the University of Bristol, UK.

    In January 2012 he was elevated to the grade of Fellow of the IEEE for his contributions to the analysis, control and applications of nonlinear systems and complex networks. On 28th February 2007 he was bestowed the title of "Cavaliere" of the Order of Merit of the Italian Republic for scientific merits from the President of Italy. In 2009, He was elected President of the Italian Society for Chaos and Complexity for the term 2010-2013. He was re-elected in 2010 for the term 2014-2017. In 2006 and again in 2009 he was elected to the Board of Governors of the IEEE Circuits and Systems Society. From 2011 to 2014 he was Vice President for Financial Activities of the IEEE Circuits and Systems Society.  In 2014 he was appointed as a member of the Board of Governors of the IEEE Control Systems Society. He was Distinguished Lecturer of the IEEE Circuits and Systems Society for the two-year term 2016-2017. His research interests include the analysis, synchronization and control of complex network systems; piecewise-smooth dynamical systems; nonlinear dynamics and nonlinear control with applications to engineering and computational biology. 
    
    He authored or co-authored more than 220 international scientific publications including more than 110 papers in scientific journals, a research monograph and two edited books. According to the international database SCOPUS (March 2018), his h-index is 38 and his publications received over 6000 citations by other authors (h-index = 48, citations = 11422 according to Google Scholar, March 2018). 
    
    He serves on the Editorial Board of several international scientific journals and conferences. From 1st January 2014 till 31st December 2015 he was Deputy Editor-in-Chief of the IEEE Transactions on Circuits and Systems: Regular Papers. He is Senior Editor of the IEEE Transactions on Control of Network Systems; Associate Editor of  Nonlinear Analysis: Hybrid Systems and the IEEE Control Systems Letters, the Conference Editorial Board of the IEEE Control System Society and the European Control Association (EUCA). He was Associate Editor of the IEEE Transactions on Circuits and Sytems I: Regular Papers from 1999 to 2002 and again from 2008 to 2010, and the IEEE Transactions on Circuits and Systems II: Brief papers from 2003 till 2008. He is regularly invited as Plenary Speaker in Italy and abroad. He received funding for over 8M Euros from research councils, the European Union and industry.
\end{IEEEbiography}

\end{document}